\documentclass[a4paper,11pt]{article}

\pdfoutput=1 

\usepackage{jheppub} 

\usepackage{wrapfig}
\usepackage{dsfont}
\usepackage{cancel}
\usepackage[svgnames,x11names]{xcolor} 
\newcommand{\be}{\begin{equation}}
\newcommand{\ee}{\end{equation}}
\newcommand{\Pexp}{{\mathrm{P}\!\exp}}
\newcommand{\nn}{\nonumber}
  
\newcommand{\sign}{\mathrm{sign}}

\newcommand{\slsh}[1]{{#1}\!\!\!\! / \, }
\usepackage{cancel}
\title{Perturbative evaluation of circular 1/2 BPS Wilson loops in ${\cal N}=6$  Super Chern-Simons theories}

\author[a]{Luca Griguolo,}
\author[b]{Gabriele Martelloni,}
\author[b]{Matteo Poggi}
\author[b]{and Domenico Seminara}

\affiliation[a]{Dipartimento di Fisica e Scienze della Terra, Universit\`a di Parma and INFN Gruppo Collegato di Parma, Viale G.P. Usberti 7/A, 43100 Parma, Italy}
\affiliation[b]{Dipartimento di Fisica, Universit\`a di Firenze and INFN Sezione di Firenze, Via G. Sansone 1, 50019 Sesto Fiorentino, Italy}

\emailAdd{griguolo@fis.unipr.it}
\emailAdd{martelloni@fi.infn.it}
\emailAdd{matteo.poggi.fi@gmail.com}
\emailAdd{seminara@fi.infn.it}

\abstract{We present a complete two-loop analysis of the quantum expectation value for circular BPS Wilson loops in ABJ(M) theories. We examine in details the 1/2 BPS case, that requires non-trivial fermionic couplings with the contour, finding perfect agreement with the exact matrix model answer at zero framing. The result is obtained through a careful application of DRED regularization scheme, combined with a judicious rearrangement of the relevant perturbative contributions that reduces the computation to simple integrals. We carefully analyze the contribution of fermions that is crucial for the consistency with the localization procedure and point out the arising of pivotal evanescent terms, discussing their meaning in relation to Ward identities.} 

\begin{document} 
\maketitle
\flushbottom
\section{Introduction}
The duality between string theory on $AdS_4 \times CP^3$ background and the ${N\cal} = 6$ superconformal Chern-Simons theory with matter, the celebrated ABJ(M) theory constructed in \cite{Aharony:2008ug,Aharony:2008gk}, represents an interesting example of the AdS/CFT correspondence. More precisely type IIA string theory on $AdS_4 \times CP^3$ is dual to the ${\cal N}=6$ Chern-Simons theory coupled to bifundamental matter, with gauge group $U(N) \times U(M)$, in the limit of large $N$, $M$ and large Chern-Simons level $k$, with 't Hooft couplings $\lambda=N/k$ and $\hat\lambda=M/k$.

An impressive amount of investigations has been devoted, in the last few years, to this novel realization of the AdS/CFT correspondence, and different observables of the theory have been carefully examined. Notably, Wilson loops operators were studied in a certain details, both for their relation with scattering amplitudes \cite{Henn:2010ps,Chen:2011vv,Bianchi:2011dg,Wiegandt:2011uu,Bianchi:2013pva,Bianchi:2013iha} and BPS avatars \cite{Drukker:2008zx,Chen:2008bp,Rey:2008bh}. 

BPS Wilson loops are well-studied objects in ${\cal N}=4$ Superconformal Yang-Mills theory in four dimensions and appeared since the early days of AdS/CFT correspondence \cite{Rey:1998ik,Maldacena:1998im}, being the gauge theory duals to fundamental string states. The most famous example, the circular 1/2 BPS Wilson-Maldacena loop, is one of the first non-trivial observables that interpolates smoothly between weak and strong coupling \cite{Erickson:2000af,Drukker:2000rr,Pestun:2007rz}, representing an highly sophisticated check of the string/gauge duality. More general families of loops, with less degree of supersymmetry, were later discovered \cite{Drukker:2007qr} and shown to be exactly computable through matrix integrals \cite{Bassetto:2008yf,Young:2008ed}.  Notably these results can be also extended to their correlators \cite{Bassetto:2009rt,Giombi:2012ep}.  A complete classification of BPS loops in this context has been finally presented in \cite{Dymarsky:2009si,Cardinali:2012sy}, 

The possibility to obtain all-order quantum results, for some class of these observables, relies on path-integral localization \cite{Pestun:2007rz}, a powerful mathematical technique reducing exact computations to saddle-point approximations.  Recently BPS Wilson loops played also a central role in deriving a nonperturbative expression for a non-BPS observable, the ${\cal N}=4$ Bremsstrahlung function \cite{Correa:2012at}, that can be also obtained applying the machinery of integrability \cite{Gromov:2012eu,Gromov:2013qga}. The program of connecting localization results with integrability computations, vigorously advocated in \cite{Drukker:2011za}, culminated instead with the discovery of a general TBA equation for  the generalized quark-anti-quark potential \cite{Drukker:2012de,Correa:2012hh}.  

In the three-dimensional maximally superconformal case, due to the presence of a Chern-Simons term, Wilson loops are the natural observables to be considered. Supersymmetric circular Wilson loops in ABJM theory were indeed firstly discussed in \cite{Drukker:2008zx}, where operators preserving 1/6 of the original supersymmetry were built as the holonomy of a generalized bosonic connection:
these operators are directly related to their four-dimensional analogue through dimensional reduction, and they also exist in Chern-Simons theories with less supersymmetry.  The quiver structure of ABJM theory leaves a certain freedom in considering linear combinations of loops transforming oppositely under time-reversal and perturbative computations for their expectation values were performed. While these operators cannot be interpretated as the gauge dual of the fundamental strings, due to the lack of supersymmetry, their exact expectation value was derived using localization techniques \cite{Kapustin:2009kz}. The computation still reduces to a matrix integral: the related matrix model was brilliantly investigated in \cite{Marino:2009jd,Drukker:2010nc,Drukker:2011zy} to obtain exact results in the large $N$ limit,  producing a non-trivial interpolating function between the weak and the strong coupling regimes of the theory. 

The gauge theory partner of the fundamental string was parallely discovered in \cite{Drukker:2009hy} and preserves 1/2 of the original supersymmetry (see \cite{Berenstein:2008dc,Lee:2010hk} for an alternative construction using the low-energy dynamics of heavy {\it W-bosons}): the contour couples not only to the gauge and scalar fields,  but also to the fermions in the bi-fundamental representation of the $U(N)\times U(M)$ gauge group. This construction can be interpretated in terms of a superconnection whose holonomy gives the 1/2 BPS Wilson loop, which turns out to be defined for any representation of the supergroup $U (N |M )$. The supersymmetry is non-trivially realized as a super-gauge transformation, exploiting therefore the full non-linear structure of the path-ordering. Remarkably, fermionic 1/2 BPS and purely bosonic 1/6 BPS loops belong to the same cohomology class, differing by a BRST exact term with respect the localization complex: their quantum expectation value should be therefore related through an appropriate linear combination of operators in the factor gauge groups.
Recently there have been a lot of advances in studying the exact expression of ABJ(M) partition function and 1/2 BPS Wilson loop at nonperturbative level \cite{Marino:2012az,Klemm:2012ii,Hatsuda:2012hm,Hatsuda:2013yua} (notably through the beautiful interpretation of ABJ(M) on $S^3$ as a Fermi gas system \cite{Marino:2011eh}). On the other hand the perturbative description and the comparison of exact results with the familiar Feynman diagrams expansion have been much less explored.  We feel instead that this issue is worth of investigation for a number of reasons, apart representing a relevant check of the localization analysis (test that has been positively performed \cite{Bianchi:2013zda}). 

A first point to understand is the mechanism of the divergences cancellation and the organization of the perturbative series for the 1/2 BPS Wilson loops. In four dimensional $N=4$ SYM it is well known that, in Feynman gauge, only exchange diagrams contributes \cite{Erickson:2000af}, reproducing almost trivially the matrix model expansion. In ABJ(M) theory, for the 1/6 BPS bosonic circle, this is not longer true \cite{Drukker:2008zx} and interacting diagrams actively partecipate. In the 1/2 BPS case we have in addition spinorial couplings and fermionic diagrams  to be properly taken into account. The divergencies associated to the exchange of fermionic propagators and to gluon-fermion vertices have been carefully studied in \cite{Griguolo:2012iq}, where a cusped loops, formed by 1/2 BPS lines, has been considered: we observed there a delicate interplay between the divergencies coming from different sectors to produce a consistent quark-antiquark potential and an exponentiated cusp anomaly. Moreover a peculiar renormalization prescription was necessary, even in some BPS limits. It is therefore important to understand the cancellation of fermionic divergencies in a ``bona fide'' BPS situation and explore the organization of the perturbative expansion for a 1/2 BPS observable. Secondly, and more crucially, we would like to address the question of the equivalence between the 1/2 BPS circular Wilson loop and the 1/6 BPS bosonic loop. A peculiar feature of the exact computations presented in \cite{Drukker:2010nc}, where both kinds of loops have been evaluated as matrix model averages, is the appearing of a preferred framing \cite{Witten:1988hf}: the final results are directly produced at framing one, without any explicit choice in the regularization procedure. Moreover the relation between the two quantum expectation values is the one derived from the claimed cohomological equivalence \cite{Drukker:2009hy}. The 1/2 BPS Wilson loop at framing one is the sum of two 1/6 BPS Wilson loops, with the trace taken alternatively in the $U(N)$ and $U(M)$ subgroup, both evaluated at framing one.  As remarked in \cite{Kapustin:2009kz}, this strongly suggests that a quantum computation, preserving the supersymmetric character of the loop observables, should be performed at framing one\footnote{ A natural choice of contours on which a circular Wilson loop could be computed at framing one is the family of Hopf fibers on $S^3$. The couplings can be chosen, in this case, so that the splitting is supersymmetry preserving. We thank Diego Trancanelli for an extensive discussion of this point.} and, consequently, the cohomological equivalence should hold in this case. A conventional perturbative computation, adopting DRED regularization \cite{Siegel:1979wq}, is instead expected to provide the Wilson loop at framing zero and therefore the result should be compared with the 1/2 BPS expression of \cite{Drukker:2010nc}, stripping out the framing phase (the perturbative computation of the 1/6 BPS loop \cite{Drukker:2008zx} indeed coincides with the framing zero expression derived from the matrix model). One immediately realizes, comparing the results of \cite{Drukker:2010nc}, that the 1/2 BPS expectation value at framing zero IS NOT the sum of two 1/6 BPS loop at framing zero! This means that the cohomological equivalence should be violated in the perturbative calculations: the diagrams depending on the fermionic couplings should contribute in a decisive way to recover the correct, non-perturbative result. This was checked in \cite{Bianchi:2013zda}, where the final expression for the relevant fermionic graphs has been reported. We should observe therefore a sort of anomaly, affecting some supersymmetric Ward identities, in the computation of the 1/2 BPS Wilson loop. As we will explicitly show, this effect appears due to the presence of evanescent terms produced when DRED regularization is carefully applied.

We present here a detailed computation, at the second non-trivial order in the perturbative expansion, of the circular 1/2 BPS Wilson loop. Instead of performing a ``brute force" calculation we preferred, in order to elucidate the origin of the anomalous contributions, to take a different approach, trying to provide a bridge with supersymmetric Ward identities. In the case of the fermionic double-exchange diagrams we succeed in analyzing in general terms the evanescent contributions, violating the cohomological equivalence, and easily recovering the result presented in \cite{Bianchi:2013zda}. The analysis of the gauge-fermions vertex is instead done in spirit of the subtraction procedure developed in \cite{Bassetto:2008yf}, for the quantum computation of general BPS loops on $S^2$, in four dimensional ${\cal N}=4$ super Yang-Mills theory. There we have been able to provide directly finite expression for this family of loops, without performing any explicit calculation in $2\omega$ dimensions, simply adding and subtracting a clever "pure gauge" contribution suggested by light-cone gauge, that takes care of the other divergent diagrams (see \cite{Bassetto:2008yf} for details). Here we mimics that technique and we are able to guess a similiar term, that allows us to isolate a crucial evanescent factor. The final result is again easily recovered in a very compact form, in terms of (almost) elementary integrals and, happily, coincides with \cite{Bianchi:2013zda}, confirming the zero framing expression of \cite{Drukker:2010nc}.

The structure of the paper is the following: in sect. 2 we discuss the general structure of 1/2 BPS circular loop and report the known results derived through localization, discussing their framing dependence. Sect. 3 is devoted to the actual perturbative computation of the 1/2 BPS case: we present first the one-loop fermionic exchange, introducing some technique that will employed in the two-loop calculation.  Then we display details our result for the fermionic double exchange diagrams and we show the appearance of the evanescent term as a violation of the supersymmetric Ward identities. Finally we apply our subtraction machinery to the diagrams involving the gauge-fermion vertex, deriving quite straightforwardly the crucial contribution that reconcile the perturbative calculation with the matrix model result at framing zero. In sect. 4 we present the complete result and in sect. 5 we draw some conclusions and discuss the perspective for future works. Appendices are instead devoted to some technical details.

{\bf Note added }: A  detailed two-loop  analysis  by means of different techniques, confirming  the results reported in  \cite {Bianchi:2013zda},  was also  presented in   \cite{Penati2}, which has  appeared concurrently in the arXiv.

\section{General features of BPS Wilson loops in ABJ(M) theories}

The construction of the 1/2 BPS Wilson loops in ABJ(M) theories mainly relies on replacing the natural $U(N)\times U(M)$  gauge  connection (represented by the two Chern-Simons gauge fields $A_\mu, \hat{A}_\mu$ of level $k$ and -$k$ respectively) with a super-connection \cite{Drukker:2009hy} 
\be
\label{superconnection}
 \mathcal{L}(\tau) \equiv -i \begin{pmatrix}
i\mathcal{A}
&\sqrt{\frac{2\pi}{k}}  |\dot x | \eta_{I}\bar\psi^{I}\\
\sqrt{\frac{2\pi}{k}}   |\dot x | \psi_{I}\bar{\eta}^{I} &
i\hat{\mathcal{A}}
\end{pmatrix} \ \  \ \ \mathrm{with}\ \ \ \  \left\{\begin{matrix} \mathcal{A}\equiv A_{\mu} \dot x^{\mu}-\frac{2 \pi i}{k} |\dot x| M_{J}^{\ \ I} C_{I}\bar C^{J}\\
\\
\hat{\mathcal{A}}\equiv\hat  A_{\mu} \dot x^{\mu}-\frac{2 \pi i}{k} |\dot x| \hat M_{J}^{\ \ I} \bar C^{J} C_{I},
\end{matrix}\ \right.
\ee
belonging to the super-algebra of $U(N|M)$. The coordinates $x^{\mu}(\tau)$ define the contour of the loop operator,  while  $M_{J}^{\ \ I}(\tau)$, $\hat M_{J}^{\ \ I}(\tau)$, $\eta_{I}^{\alpha}(\tau)$ and $\bar{\eta}^{I}_{\alpha}(\tau)$ describe the effective couplings of the scalar $C_I,\bar{C}^I$ and of the fermions $\psi_I,\bar{\psi}^I$ with the circuit. The fermionic couplings, in particular, are Grassmann even quantities even though they transform in the spinor representation of the Lorentz group.

The  free  parameters appearing in \eqref{superconnection} can be constrained by imposing that the Wilson loop resulting from the superconnection is globally supersymmetric. This issue is delicate: the familiar requirement $\delta_{\rm susy}\mathcal{L}(\tau)=0$ does not  yield  any 1/2 BPS solution indeed. One  just obtains  loop operators which are merely bosonic ($\eta=\bar\eta=0$) and  at most $1/6$ BPS \cite{Drukker:2008zx,Rey:2008bh}. A clever way to get 1/2 BPS solution consists in replacing  $\delta_{\rm susy}\mathcal{L}(\tau)=0$ with  the weaker condition \cite{Drukker:2009hy,Lee:2010hk}
\be
\label{var1}
\delta_{\rm susy}\mathcal{L}(\tau)=\mathfrak{D}_{\tau} G\equiv\partial_{\tau} G+ i\{ \mathcal{L},G],
\ee
where  the r.h.s. is the super-covariant derivative  constructed out of the connection  
$\mathcal{L}(\tau)$  acting on a super-matrix $G$ in $\mathfrak{u}(N|M)$. The requirement \eqref{var1} 
assures that the action of the relevant  supersymmetric generators translates into an infinitesimal
 super-{\it gauge} transformation for $\mathcal{L}(\tau)$ and thus   the {\it traced} loop operator is invariant. 

An explicit solution of the 1/2 BPS case has been constructed in \cite{Drukker:2009hy} when the contour is a straight line or a circle. In the circular case the matrices $M_{J}^{\ \ I}(\tau)$ and $\hat M_{J}^{\ \ I}(\tau)$ tuning the interactions  with the scalar bilinears are diagonal and constant and they are simply given by $M_{J}^{\ \ I}=\hat M_{J}^{\ \ I}=\mathrm{diag}(-1,1,1,1)$. The fermionic couplings $\eta_{I}(\tau)$ and $\bar \eta^{I}(\tau)$ have a factorized  structure and they can be written  as
\be
\label{coup}
\eta_{I}^{\alpha}= n_{I}\eta_{\alpha}=(e^{\frac{i\tau}{2}}\ \ -i e^{-\frac{i\tau}{2}})\begin{pmatrix}1 \\ 0 \\ 0 \\ 0\end{pmatrix}\ \ \ \ \ 
\bar\eta^{I}_{\alpha}= \bar n^{I}\bar\eta_{\alpha}=(1\ 0 \ 0  \ 0)\begin{pmatrix} i e^{-\frac{i\tau}{2}} \\ -   e^{\frac{i\tau}{2}}\end{pmatrix},
\ee
where we have used the usual parametrization for the unit circle
\be
x^{1}=\cos\tau,\ \ \  x^{2}=\sin\tau,\ \ \  x^{3}=0.
\ee
In terms of the superconnection $\mathcal{L}(\tau)$, the Wilson loop can be defined as the anti-path-ordered exponential
\be
\mathcal{W}_{\cal R}=\frac{1}{{\rm dim}_{\cal R}}\mathrm{Tr}_{\cal R}\left[\Pexp\left(i\oint_{0}^{2\pi}  d\tau \mathcal{L}(\tau)\right)\right], 
\ee
where ${\cal R}$ is a representation of $U(N|M)$. We remark that in order to obtain an invariant result it is mandatory to take the trace of the superholonomy and not, as naively it could be expected, the supertrace (this is related to the anti-periodicity condition obeyed by fermionic couplings along the circle  \cite{Drukker:2009hy}). As we have anticipated, a purely bosonic 1/6 BPS circle is instead obtained by choosing $\eta_{I}(\tau)=\bar \eta^{I}(\tau)=0$: it requires, in turn, a different choice of the bosonic matrices $i.e$ $M_{J}^{\ \ I}=\hat M_{J}^{\ \ I}=\mathrm{diag}(-1,-1,1,1)$. This case was originally studied in \cite{Drukker:2008zx}, both at weak and strong coupling , and it was observed there that it could not be considered the QFT dual of the fundamental string in $AdS^4\times CP_3$. Later on it was computed exactly \cite{Kapustin:2009kz} through localization. In spite of the different BPS degree, the 1/2 BPS circle is cohomologically equivalent at classical level to its 1/6 BPS counterpart \cite{Drukker:2009hy}. The key point, in order to establish the equivalence of the two observables, was to notice that the difference between $\mathcal{W}_{\cal R}^{1/2}$ and  $\mathcal{W}_{\cal R}^{1/6}$ can be cast into a $Q$-exact term
\be
\label{Qex}
 {\rm dim}_{\cal R}\,\mathcal{W}_{\cal R}^{1/2}-\left({\rm dim}_{{\cal R }_N}\mathcal{W}_{{\cal R}_N}^{1/6}+{\rm dim}_{{\cal R }_M}\mathcal{W}_{{\cal R}_M}^{1/6}\right)=QV.
\ee
Here $Q$ is a particular supercharge, constructed using the fermionic and bosonic couplings, and it generates transformations leaving invariant both operators. Its explicit expression and the precise form of $V$ are reported in \cite{Drukker:2009hy}. Because $Q$ can be chosen to generate also the BRST complex used in the localization procedure, we would expect that the quantum expectation value of 1/2 BPS and 1/6 BPS Wilson loops are basically the same.
\be
\label{Qex2}
 \langle\mathcal{W}_{\cal R}^{1/2}\rangle=\frac{1}{{\rm dim}_{\cal R}\,}\left({\rm dim}_{{\cal R}_N}\langle\mathcal{W}_{{\cal R}_N}^{1/6}\rangle+{\rm dim}_{{\cal R}_M}\langle\mathcal{W}_{{\cal R}_M}^{1/6}\rangle\right).
\ee
Actually the presence of quantum infinities needs a regularization procedure, that could potentially affect the classical cohomological equivalence: an explicit example in the ABJ(M) case was observed in \cite{Griguolo:2012iq}, where the divergencies structure of the 1/2 BPS line has been found different from the one of the related 1/6 BPS line. In the present situation this issue is rather subtle because we expect that BPS observables should be finite and regularization is needed only in the intermediate step. On the other hand it is well known that in pure Chern-Simons theory Wilson loops depend in a very specific way from a regularization choice, the so called ${\it framing}$ \cite{Witten:1988hf}, and a global phase appears in the quantum evaluation, parameterizing the different possibilities. The non-perturbative evaluation of the BPS Wilson loops in ABJ(M) seems to display a similar phenomenon. 

By means of localization techniques, the path integral of the ${\cal N}=6$ superconformal Chern-Simons theory on $S^3$ can be exactly written as a particular, non-Gaussian matrix model \cite{Kapustin:2009kz}. The partition function is obtained from the following matrix integral 
\begin{eqnarray}
\mathcal{Z}&=&\int \prod_{a=1}^{N}d\lambda _{a} \ e^{i\pi k\lambda
_{a}^{2}}\prod_{b=1}^{M}d\hat{\lambda }_{b} \ e^{-i\pi k\widehat{%
\lambda }_{b}^{2}} \times \label{partf}\\
&& \frac{\prod_{a<b}^{N}\sinh ^{2}(\pi (\lambda
_{a}-\lambda _{b}))\prod_{a<b}^{M}\sinh ^{2}(\pi (\hat{\lambda 
}_{a}-\hat{\lambda }_{b}))}{\prod_{a=1}^{N}\prod_{b=1}^{M}\cosh ^{2}(\pi (\lambda _{a}-\hat{\lambda }_{b}))}\nonumber 
\end{eqnarray}
The quantum expectation values of ${\cal W}^{1/6}_{{\cal R}_N}$ and $\hat{{\cal W}}
^{1/6}_{{\cal R}_M}$, in the fundamental representations, are directly obtained by inserting in (\ref{partf}) the functions
\be
w^{1/6}_N=\frac{1}{N}\sum_{a=1}^{N}e^{2\pi \lambda _{a}}\quad \text{%
and}\quad \hat{w}^{1/6}_M=\frac{1}{M}\sum_{a=1}^{M}e^{2\pi 
\hat{\lambda }_{a}}  \label{W1/6}
\ee
corresponding to the $U(N)$ and $U(M)$ pieces respectively. The computation of the $1/2$ BPS Wilson loop, in the fundamental representation $F$, is instead equivalent to the insertion in (\ref{partf}) of the operator \cite{Drukker:2010nc}
\be
w^{1/2}_F=\frac{1}{N+M}\left( \sum_{a=1}^{N}e^{2\pi \lambda _{a}}+\sum_{a=1}^{M}e^{2\pi \hat{\lambda }_{a}} \right). \label{W1/2}
\end{equation}
The localization procedure implies therefore the expected relation between the quantum expectation values of $1/6$ BPS and the $1/2$ BPS operators 
\be
\langle{\cal W}^{1/2}_F\rangle_{\cal Z} =\frac{N\langle
{\cal W}^{1/6}_N\rangle_{\cal Z} +M<\langle\hat{{\cal W}}^{1/6}_M\rangle_{\cal Z}}{N+M},
\label{WF}
\ee
where with $\langle\,\,\,\,\rangle_{\cal Z}$  we have denoted the quantum avarages obtained from the matrix model eq. (\ref{partf}). In deriving eq. (\ref{WF}) it has been tacitly assumed that the regularization procedure preserves the cohomological relation: it is therefore tempting to analyze the framing dependence of this result. The 1/6 BPS case was discussed at perturbative level in \cite{Drukker:2008zx}, employing conventional DRED regularization at framing $f=0$: comparing the explicit two-loop expression with the expansion of the matrix model average at the same order, one discovers that
\begin{align}
\label{fram1/6}
& \langle{\cal W}_N^{1/6} \rangle_{\cal Z} = e^{\frac{i\pi}{k}N} \, \langle{\cal W}^{1/6}_N\rangle^{f=0} \notag \\
& \langle\hat{{\cal W}}^{1/6}_M\rangle_{\cal Z}  = e^{-\frac{i\pi}{k}M} \, \langle\hat{{\cal W}}_M^{1/6}\rangle^{f=0} .
\end{align}
Localization computes the Wilson loop at framing $f=1$: this feature was argued in \cite{Kapustin:2009kz} by studying in this framework the circular loop in pure Chern-Simons theory on $S^3$. For the $1/2$ BPS observable, the framing factor is immediately identified by direct comparison with the matrix model average \cite{Drukker:2010nc} 
\begin{align}
\label{fram1/2}
& \langle{\cal W}^{1/2}_F\rangle_{\cal Z} = e^{\frac{i\pi }{k}(N-M)} \langle{\cal W}_F^{1/2}\rangle^{f=0} =   e^{\frac{i\pi }{k}(N-M)}  \notag \\
& \times \left[ 1 -\frac{\pi^2}{6\, k^2} \left( N^2+M^2 - 4 NM \right)+ \mathcal{O}(1/k^3) \right].
\end{align}
The square bracket should represent the perturbative result at framing zero, that we would better recover in the next sections. We remark that the above relations impliy that the 1/2 BPS Wilson loop at framing $f=0$ is not given, at quantum level, by the sum of the two (bosonic) 1/6 BPS Wilson loops. This means, in particular, that fermionic interactions should play a crucial role, at perturbative level, to find agreement with the localization procedure.

\section{Perturbative evaluation}
\subsection{Generalities}
Before describing the details of the  two-loop perturbative evaluation  of the circular Wilson loop, we  shall briefly summarize the general framework for our analysis. The quantum holonomy of the super-connection ${\cal L}$  in a representation ${\cal R}$ of the supergroup $U(N|M)$ is by definition
\be
\label{eq:loopexpectationvalue}
\left\langle \mathcal{W}_{\cal R} \right\rangle= \frac{1}{{\rm dim}_{\cal R}}\int {\cal D}[A,\hat{A},C,\bar{C},\psi,\bar{\psi}]~
{\rm e}^{-S_{\rm ABJ(M)}}~{\rm Tr}_{\cal R} \left[
  {\rm P} \exp \left(i\oint_{C} d\tau\, {\cal L}(\tau) \right) \right],
\ee
where $S_{ABJ(M)}$ stands for the action for $ABJ(M)$ theories in euclidean space.  The part relevant for us  is presented in app. \ref{ABJ}.  In the following $\mathcal{R}$ is taken to be the fundamental representation and $\mathcal{C}$ to be the circle of unit radius
in the plane $x_{3}=0$. 

To begin with,  we  shall only consider    the upper left $N\times N$ block of the super-matrix appearing  in \eqref{eq:loopexpectationvalue}.
 For this sector the trace  in \eqref{eq:loopexpectationvalue} is obviously taken in the fundamental representation ${\bf N}$ of  $U(N)$. The expectation value  of  the lower 
 diagonal  block can be then obtained  from the above analysis by replacing $N$ with $M$. A two-loop computation   requires to expand the path-exponential   in \eqref{eq:loopexpectationvalue} up to  the fourth order.  The expansion of  the upper block at this order will  include both contributions of  {\it bosonic} and {\it fermionic} type:
\begin{align}
\label{expaloop}
\mathbb{W}_{\mathbf{N}}&={\rm Tr}_{\mathbf{N}}\left[1+i\int_\Gamma d\tau_1{\cal A}_1-\int_{\Gamma}d\tau_{\mbox{\tiny $\displaystyle1\!\!>\!\! 2$}}\Biggl({\cal A}_1{\cal A}_2-(\eta\bar{\psi})_1(\psi\bar{\eta})_2 \Biggr) \right.\nonumber\\
&-i\int_{\Gamma}d\tau_{\mbox{\tiny $\displaystyle1\!\!>\!\! 2\!\!>\!\!3$}}\Biggl( {\cal A}_1{\cal A}_2{\cal A}_3+\frac{2\pi}{k}[(\eta\bar{\psi})_1(\psi\bar{\eta})_2{\cal A}_3 
+(\eta\bar{\psi})_1\hat{\cal A}_2 (\psi\bar{\eta})_3+{\cal A}_1(\eta\bar{\psi})_2(\psi\bar{\eta})_3]\Biggr)\nonumber\\
&\left.+\int_{\Gamma}d\tau_{\mbox{\tiny $\displaystyle1\!\!>\!\! 2\!\!>3\!\!>4$}}\left(\left(\frac{2\pi}{\kappa}\right)^{2}(\eta\bar{\psi})_1(\psi\bar{\eta})_2(\eta\bar{\psi})_3(\psi\bar{\eta})_4+\mathcal{A}_{1}\mathcal{A}_{2}\mathcal{A}_{3}\mathcal{A}_{4}-\right.\right.\\ 
&-\left(\frac{2\pi}{\kappa}\right)\mathcal{A}_{1}\mathcal{A}_{2}(\eta\bar{\psi})_3(\psi\bar{\eta})_4-\left(\frac{2\pi}{\kappa}\right)\mathcal{A}_{1}(\eta\bar{\psi})_2\hat{\mathcal{A}}_{3}(\psi\bar{\eta})_4-\left(\frac{2\pi}{\kappa}\right)(\eta\bar{\psi})_1\hat {\mathcal{A}_{2}}\hat{\mathcal{A}_{3}}(\psi\bar{\eta})_4-\nonumber\\
&\left.\left.
-\left(\frac{2\pi}{\kappa}\right)\mathcal{A}_{1}(\eta\bar{\psi})_2(\psi\bar{\eta})_3\mathcal{A}_{4}-\left(\frac{2\pi}{\kappa}\right)(\eta\bar{\psi})_1\hat{\mathcal{A}}_{2}(\psi\bar{\eta})_3\mathcal{A}_{4}-\left(\frac{2\pi}{\kappa}\right)(\eta\bar{\psi})_1(\psi\bar{\eta})_2\mathcal{A}_{3}\mathcal{A}_{4}
\right)\right].\nonumber
\end{align}
In \eqref{expaloop} we have introduced a shorthand notation for the circuit parameter dependence of the fields, namely ${\cal A}_i = {\cal A}(x_i)$ with $x_i = x(\tau_i)$. Above we have  suppressed the spinor and $SU(4)_R$ indices  [$\eta\bar{\psi}\equiv \eta_{I}^{\alpha}\bar\psi^{I}_{\alpha}$, $\psi\bar{\eta}\equiv \psi_{I}^{\alpha}\bar{\eta}^{I}_{\alpha}$] and  we have used that  $|\dot{x}|=1$ for our  parametrization.

Since we shall perform our computation at framing zero and  our contour lies on a plane, any diagram involving  a three-level gauge propagator ranging between two points of the circle yields zero. In fact one can immediately realize that this type of  graphs will always 
 \begin{wrapfigure}[10]{r}{65mm}
\begin{center} \includegraphics[width=40mm]{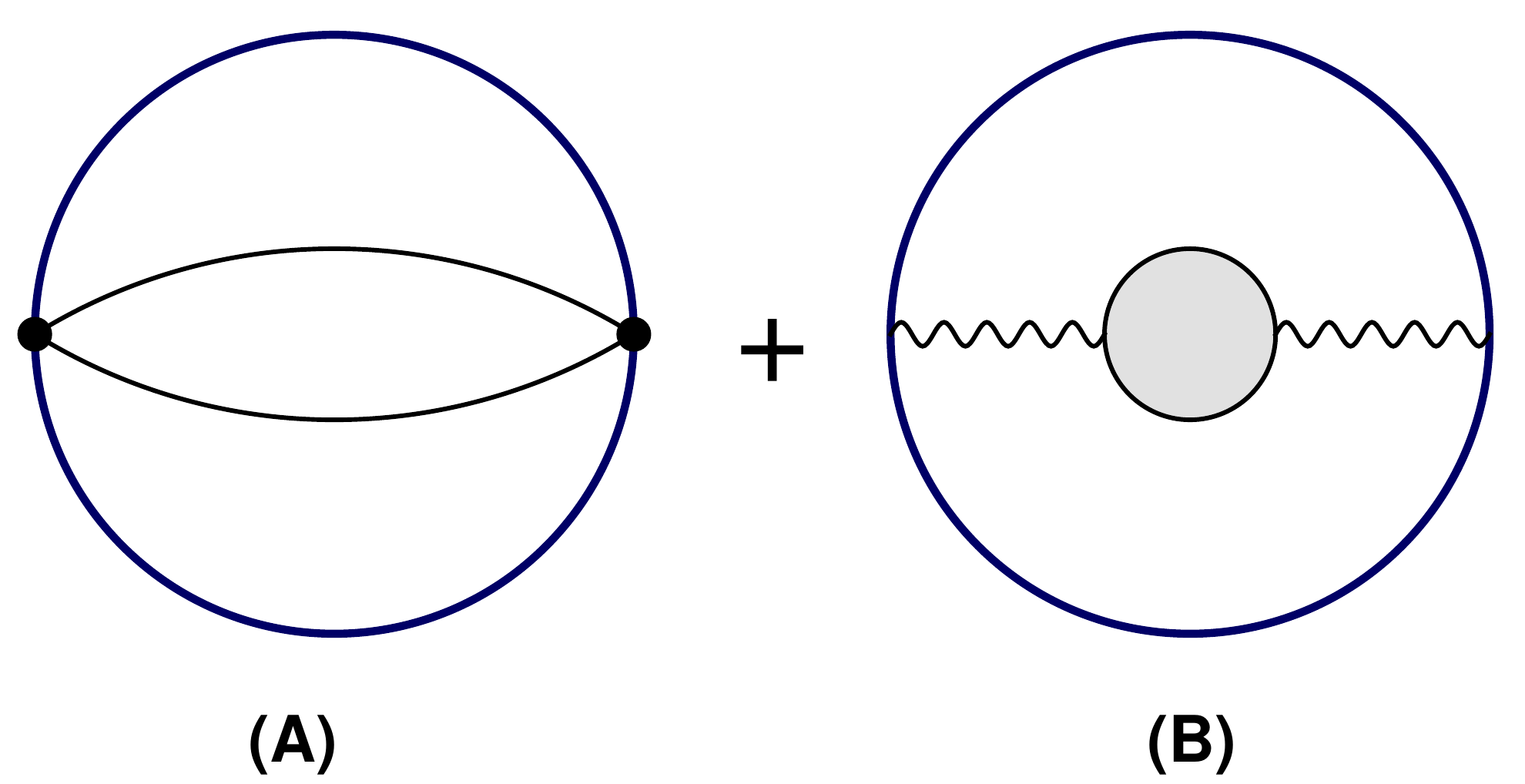}\end{center}
\vskip -6mm 
\caption{\label{twobloopdiag}  The first graph is the scalar contribution,
while the second one is the one-loop correction to the gauge propagator.}
\end{wrapfigure}
 contain a Levi-Civita tensor contracted with three linear dependent vectors.  Therefore at this order of the perturbative expansion  we can neglect all  the terms in \eqref{eq:loopexpectationvalue}
of the type $\mathcal{A}^{4}$  and $\mathcal{A}^{2}\psi^{2}$.

The integral over a single ${\cal A}$ yields  a tadpole-like scalar graph which is  zero in DRED. 
The  bosonic monomial ${\cal A}_1{\cal A}_2$ is irrelevant at one-loop. It gives origin to a diagram with a single gauge propagator connecting two points of the circle, which vanishes for the reason mentioned above.  However at two loops  this term becomes active and  produces  the two graphs in fig. \ref{twobloopdiag}. Since the constant matrix $M_{I}^{\ J}$, governing the scalar couplings, appears quadratically  in  both diagrams, they are identical to the those  computed in \cite{Drukker:2008zx} for the $1/6$ BPS circle and we can borrow their result  
\be
\mathbf{(A)}+\mathbf{(B)}=\frac{\pi^{2}}{\kappa^{2}} \frac{N^{2} M}{N+M}.
\ee
 
\begin{wrapfigure}[10]{l}{58mm}
\begin{center} \includegraphics[width=33mm]{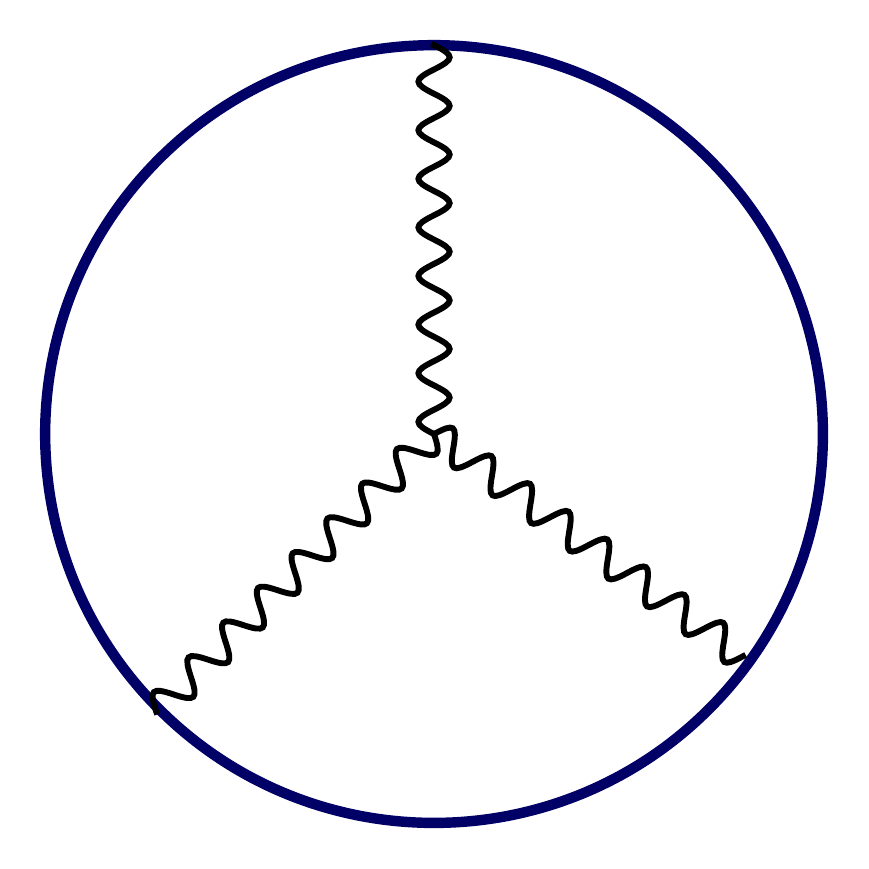}\end{center}
\vskip -7mm 
\caption{\label{GaugeVertexdiag}   \small  Gauge vertex diagram}
\end{wrapfigure}

The next step is to consider the monomial ${\cal A}_1{\cal A}_2{\cal A}_3$. At this order only the gauge field in $\mathcal{A}$ are relevant
and one finds the vertex diagram  given in  fig. \ref{GaugeVertexdiag}. 
This graph  also appears  in pure Chern-Simons theory and its value  only depends  on the topology of the loop. The circle is an {\it unknot}, for which the explicit result was originally  computed in \cite{Guadagnini:1989am}.
Translated in the language relevant for ABJ(M) Wilson loops it is given by 
\begin{equation}
-\frac{N^{3}}{N+M}\frac{\pi^{2}}{6\kappa^{2}}.
\end{equation}
The above analysis exhausts all the diagrams which are merely bosonic. It remains to compute the diagram  which involves fermions propagating along the contour. We have three type of contributions: (1) from the monomial $\bar\psi_{1}\psi_{2}$ we have the so-called single exchange diagram discussed  in subsec. \ref{single-exc}; (2) from the  four fermion terms we obtains the double-exchange diagrams whose explicit evaluation is performed in subsec.
\ref{DoubleExchange}; (3) finally we consider the contribution coming from $\psi\bar\psi A$ monomials, which will be analyzed  in subsec. \ref{Vertdiag}.  This last family of diagram hides one of the most delicate point  of this computation.

\subsection{Single Exchange Diagram}
\label{single-exc}
The expectation value of the monomial  $(\eta\bar\psi)_{1}(\psi\bar\eta)_{2}$ potentially contributes to the perturbative expansion. At the lowest order only the free Wick-contraction of the two fermonic fields    
 \begin{wrapfigure}[8]{l}{50mm}
\begin{center} 
\includegraphics[width=27mm]{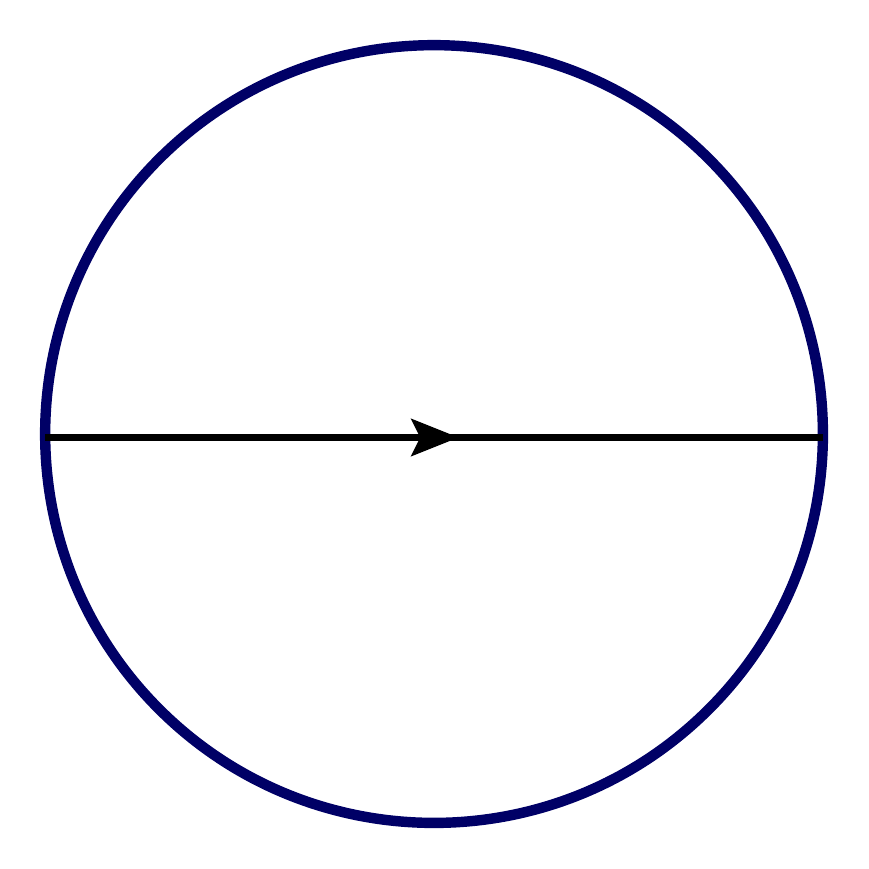}\end{center}
\vskip -7mm 
\caption{\label{1-loopdiag}   \small Fermion exchange.}
\end{wrapfigure}
appears, which  yields the diagram  schematically represented  
in fig. \ref{1-loopdiag}.  The value of this diagram is obtained by computing the contour integral 
 \begin{align}
 \label{1-loopeq1}
\left(\frac{2\pi}{\kappa} \right)\int_{0}^{2\pi}\!\!\!d\tau_{1}\int_{\tau_{1}}^{2\pi} \!\!\!d\tau_{2}~\langle\mathrm{Tr}[(\eta\bar\psi)_{1}(\psi\bar\eta)_{2}] \rangle_{0},
 \end{align}
 where $\langle\cdots\rangle_{0}$ represents the VEV in the free theory.
If we  use the explicit form of the fermion propagator given in app. \ref{ABJ}
and the explicit parametrization of the contour,  we can cast the integral as follows 
\be
 \label{1-loopeq2}
\left\langle \mathrm{Tr}[ (\eta\bar\psi)_{1}(\psi\bar\eta)_{2}]\right\rangle_{0}=\frac{M N}{4^{1-\epsilon}\pi^{\frac{3}{2}-\epsilon}} 
   \Gamma \left(\frac{3}{2}-\epsilon
   \right)\left[ \csc ^2\left(\frac{\tau _1-\tau
   _2}{2}\right)\right]^{\frac{3}{2}-\epsilon } \sin \left(\frac{\tau _1-\tau
   _2}{2}\right).
\ee 
To be in agreement  with the matrix model prediction, the path-ordered integral \eqref{1-loopeq1} of the quantity \eqref{1-loopeq2} must yield a vanishing result, when DRED is used. Before proceeding with the actual computation, we find instructive  briefly review how this result is implied by the fact that the 
loop, at this order, is still cohomologically equivalent to the $1/6$ BPS  circular operator. 
In the proof  presented in  \cite{Drukker:2009hy} the  key-ingredient is the gauge function 
\be
\Lambda=i\sqrt{\frac{\pi}{2\kappa}} e^{\frac{i\tau}{2}}\begin{pmatrix} 0 & C_{2}\\ \bar C^{2} & 0\end{pmatrix}
\ee
which transforms as follows
\be
Q\Lambda(\tau)=L_{F}(\tau),\ \ \ \ \  QL_{F}(\tau)=-8 \mathcal{D}_{\tau}(e^{-i\tau}\Lambda(\tau)),
\ee
 when acting with the relevant  supersymmetry charge $Q$. Here $L_{F}$ is the fermionic part of the super-connection $\mathcal{L}$ and the above relations only hold when the fields evaluated along the circuit.  Let us consider now  the quantity    
\be
R_{1}=\frac{1}{2} [\Lambda(\tau_{1}) L_{F}(\tau_{2})- L_{F}(\tau_{1})\Lambda(\tau_{2})],
\ee
and take its variation under the action of the supercharge $Q$:
\be
\label{QR1}
Q R_{1}= L_{F}(\tau_{1}) L_{F}(\tau_{2})-4 [\Lambda(\tau_{1}) D_{\tau_{2}}( e^{-i\tau_{2}}\Lambda(\tau_{2}))-D_{\tau_{1}}( e^{-i\tau_{1}}\Lambda(\tau_{1}))\Lambda(\tau_{2}) ].
\ee
 Under the assumptions  that 
$\langle Q(\mathrm{anything})\rangle=0$ eq. \eqref{QR1} becomes the  following Ward-identity 
\be
\label{Ward1}
\langle L_{F}(\tau_{1}) L_{F}(\tau_{2})\rangle=4[\langle\Lambda(\tau_{1}) D_{\tau_{2}}( e^{-i\tau_{2}}\Lambda(\tau_{2}))\rangle-\langle D_{\tau_{1}}( e^{-i\tau_{1}}\Lambda(\tau_{1}))\Lambda(\tau_{2})\rangle].
\ee
At  this order in perturbation theory eq. \eqref{Ward1} simply translates into a differential relation  between  the tree-level fermion and  scalar propagators attached to the contour 
\be
\label{Ward2}
 \langle(\eta\bar\psi)_{1}(\psi\bar\eta)_{2}\rangle_{0}=\Bigl[\partial_{\tau_{2}}\bigl(e^{\frac{i}{2}(\tau_{1}-\tau_{2})}\langle C_{2}(\tau_{1}) \bar C^{2}(\tau_{2})\rangle_{0}\bigr)-\partial_{\tau_{1}}\bigl(e^{-\frac{i}{2}(\tau_{1}-\tau_{2})}\langle C_{2}(\tau_{1}) \bar C^{2}(\tau_{2})\rangle_{0}\bigr)\biggr].
\ee 
One can easily check that  \eqref{Ward2} is satisfied if  the dimension is exactly three.  In particular one can formally show that the result of  the fermion  contribution cancels exactly with a tadpole-like diagram coming from the  monomial $\mathcal{A}_{1}$ in \eqref{expaloop}. This formal argument breaks down at the quantum level since we are dealing with divergent  quantities and a regularization is needed. 

Let us examine what  happens if we introduce a regularization scheme such as DRED, where   $D=3-2\epsilon$.  By means of a direct computation, we can show that the identity  \eqref{Ward2} is softly broken  by an {\it anomalous}  term proportional to $\epsilon$, leading to the following
modification
\begin{align}
\label{Ward3}
 \langle(\eta\bar\psi)_{1}(\psi\bar\eta)_{2}\rangle_{0}=&\Bigl[\partial_{\tau_{2}}\bigl(e^{\frac{i}{2}(\tau_{1}-\tau_{2})}\langle C_{2}(\tau_{1}) \bar C^{2}(\tau_{2})\rangle_{0}\bigr)-\partial_{\tau_{1}}\bigl(e^{-\frac{i}{2}(\tau_{1}-\tau_{2})}\langle C_{2}(\tau_{1}) \bar C^{2}(\tau_{2})\rangle_{0}\bigr)\biggr]-\nn\\
 &-\epsilon\frac{\Gamma
   \left(\frac{1}{2}-\epsilon \right)}{4^{1-\epsilon } \pi ^{\frac{3}{2}-\epsilon
   }}
\frac{\ \ \left[\sin
   ^2\frac{\tau _1-\tau
   _2}{2} \right]^{
   \frac{1}{2}+\epsilon}}{\sin\frac{\tau _1-\tau
   _2}{2} }.
 \end{align}
 We can now safely integrate both sides of eq. \eqref{Ward3}  along the contour: during this process we drop all the tadpole-like  contributions arising from the integration of the derivative term since they vanish  in our regularization scheme.  On the r.h.s. only the anomalous term survives and we obtain
\be
\label{Ward4}
\frac{   \Gamma \left(\frac{3}{2}-\epsilon
   \right)}{2^{2-2 \epsilon } \pi ^{\frac{3}{2}-\epsilon}} \int_{0}^{2\pi}\!\!\!\! \!d\tau_{1}\int_{\tau_{1}}^{2\pi}\!\!\!\! \!d\tau_{2}\frac{\sin \frac{\tau _1-\tau _2}{2}}{\left[\sin ^2\left(\frac{\tau _1-\tau
   _2}{2}\right)\right]^{\frac{3}{2}-\epsilon}}=\epsilon    \frac{\Gamma
   \left(\frac{1}{2}-\epsilon \right)}{4^{1-\epsilon } \pi ^{\frac{3}{2}-\epsilon
   }}\int_{0}^{2\pi}\!\!\!\! \!d\tau_{1}\int_{\tau_{1}}^{2\pi}\!\!\!\! \!d\tau_{2}\!\!\!
{\ \ \left[\sin
   ^2\frac{\tau _1-\tau
   _2}{2} \right]^{
   \epsilon}},
   \ee
 once we have used the explicit expression for the propagators. If we finally take the limit  $\epsilon\to 0$ on both sides, we find
 \be
\lim_{\epsilon\to 0}
\frac{   \Gamma \left(\frac{3}{2}-\epsilon
   \right)}{2^{2-2 \epsilon } \pi ^{\frac{3}{2}-\epsilon}} \int_{0}^{2\pi}\!\!\!\! \!d\tau_{1}\int_{\tau_{1}}^{2\pi}\!\!\!\! \!d\tau_{2}\frac{\sin \frac{\tau _1-\tau _2}{2}}{\left[\sin ^2\left(\frac{\tau _1-\tau
   _2}{2}\right)\right]^{\frac{3}{2}-\epsilon}}=0
 \ee
 since  the integral on the r.h.s of  \eqref{Ward3} is finite for $\epsilon$ 
 approaching 0.  Namely  the one-loop contribution for the fermions vanishes and the anomalous  term  is ineffective when we remove the regularization.    In the next section we shall see that this  does not  occur at 
 two-loop (un)fortunately.
 
The integral \eqref{1-loopeq1} can be of course directly computed  for any $\epsilon$. In fact it can be rearranged as follows
\begin{align}
&-\frac{2\pi}{\kappa} \frac{M N}{4^{1-\epsilon}\pi^{\frac{3}{2}-\epsilon}} 
   \Gamma \left(\frac{3}{2}-\epsilon
   \right)
\int_{0}^{2\pi}\!\!\!d\tau_{1}\int_{\tau_{1}}^{2\pi} \!\!\!d\tau_{2}~\left[ \csc ^2\left(\frac{\tau _2-\tau_1}{2}\right)\right]^{1-\epsilon }\!\!\! 
=\nonumber\\
=&-\frac{\pi}{\kappa} \frac{M N}{4^{1-\epsilon}\pi^{\frac{3}{2}-\epsilon}} 
   \Gamma \left(\frac{3}{2}-\epsilon
   \right)
\int_{0}^{2\pi}\!\!\!d\tau_{1}\int_{0}^{2\pi} \!\!\!d\tau_{2}~\left[ \csc ^2\left(\frac{\tau _2-\tau_1}{2}\right)\right]^{1-\epsilon }\!\!\! 
=\nonumber\\
=&-\frac{2\pi^{2}}{\kappa} \frac{M N}{4^{1-\epsilon}\pi^{\frac{3}{2}-\epsilon}} 
   \Gamma \left(\frac{3}{2}-\epsilon
   \right)
\int_{0}^{2\pi}\!\!\!d\tau~\left[ \csc ^2\left(\frac{\tau}{2}\right)\right]^{1-\epsilon }\!\!\! 
=
   \frac{M N 4^{\epsilon } \pi ^{\epsilon
   +2} \sec (\pi  \epsilon )}{\kappa 
   \Gamma (\epsilon )},
\end{align}
where we have used the fact that the integrand has become symmetric to close  the region of integration.
For $\epsilon$ approaching zero, we  again get a vanishing result!

At two-loop the monomial $(\eta\bar\psi)_{1}(\psi\bar\eta)_{2}$ produces
a diagram involving  the one-loop corrected fermion propagator. However this contribution being proportional to $N-M$ exactly cancels when we sum
upper and lower blocks.
\subsection{Fermionic double exchange diagrams}
  \label{DoubleExchange}

 We come now to discuss a more subtle  group of  diagrams, namely those involving
 \begin{wrapfigure}[9]{l}{78mm}
\centering{
    \includegraphics[width=.4\textwidth, height=.2\textwidth]{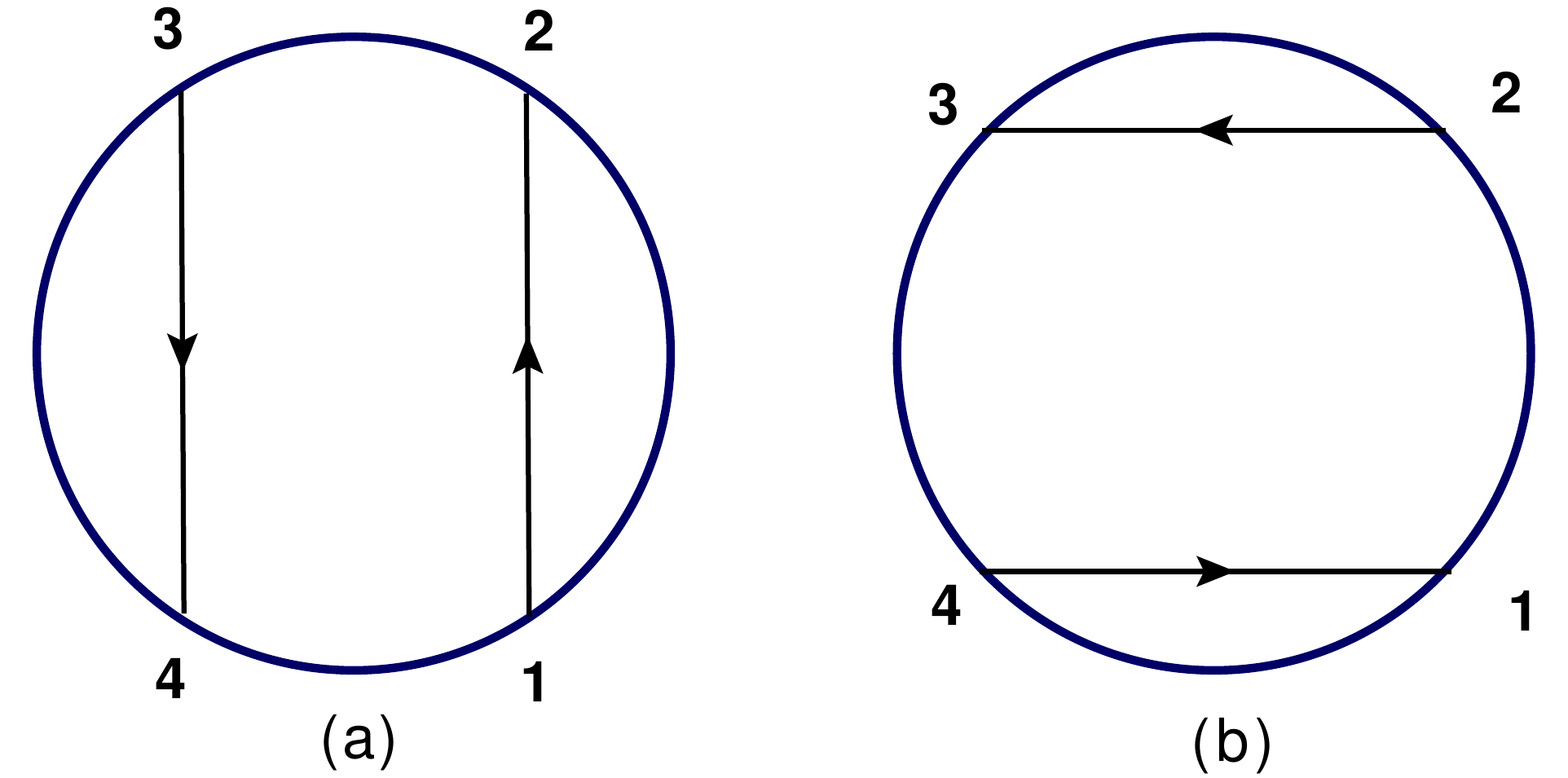}
   }
\vskip -.3cm
\caption{\label{DExa} Fermionic double exchange diagrams.}
\end{wrapfigure}  
  two $\langle\psi\bar\psi\rangle$ propagators. They arise when we evaluate the Wick-contractions of the  fermionic quadrilinear present  in \eqref{expaloop}. At this order in perturbation theory,  the expansion of the term $(\bar\psi\psi\bar\psi\psi)$, present in the  upper block, gives origin only to two sets of  non-vanishing contractions, weighted  by different group factors.  Diagramatically the result is depicted in fig. \ref{DExa}. We remark the absence  of  a crossed (non planar) diagram which would be present in the case of gauge fields. Its absence is a consequence  of dealing with fields which carry a $U(1)$ charge.

The contour integral describing the contribution in fig.  \ref{DExa} is given by
\begin{align}
\label{DEX}
-\left(\frac{2\pi}{\kappa}\right)^{2}\!\!\!\oint_{C}\!\!d\tau_{\mbox{\tiny $\displaystyle4\!\!>\!\! 3\!\!>\!\!
2 \!\!>\!\! 1$}} \mbox{\small $\displaystyle \left[
M^{2}N
\underset{\mathbf{(I)}}{\langle (\bar\eta\psi)_{2} (\bar\psi\eta)_{1})\rangle_{0}\langle(\psi\bar\eta)_{4}(\eta\bar\psi)_{3}\rangle_{0}}- N^{2} M\underset{\mathbf{(II)}}{\langle(\psi\bar\eta)_{2}(\eta\bar\psi)_{3}\rangle_{0}\langle(\psi\bar\eta)_{4} (\eta\bar\psi)_{1}\rangle_{0}}\right],$}
\end{align}
where we have already performed the trace over the gauge index in order to stress the different dependence on $M$ and $N$ of the two terms. If we use the explicit expression of the fermion propagator given in app.  $\ref{ABJ}$ and that  of the fermionic couplings we obtain the following two integrals to evaluate
\begin{align}
\!\!\mathrm{(\mathbf{I})}=&
-\frac{\Gamma^{2}
   \left(\frac{3}{2}-\epsilon
   \right)}{4^{2-2 \epsilon } \pi ^{3-2
   \epsilon}  }\int_{0}^{2\pi}\!\!\!\! d\tau_{1}\int_{\tau_{1}}^{2\pi} \!\!\!\!d\tau_{2}\int_{\tau_{2}}^{2\pi} \!\!\!\!d\tau_{3}\int_{\tau_{3}}^{2\pi} \!\!\!\!d\tau_{4}~
 \frac{ \sin\left(\frac{\tau _1-\tau_2}{2}
   \right) \sin
   \left(\frac{\tau _3-\tau_4}{2}\right) }{\left[\sin
   ^2\left(\frac{\tau _1-\tau_2}{2}
  \right){}\right]^{ \frac{3}{2}-\epsilon} \left[\sin
   ^2\left(\frac{\tau _3-\tau_4}{2}
   \right){}\right]^{\frac{3}{2}-\epsilon }}=\nonumber\\
   =&
-\frac{\Gamma^{2}
   \left(\frac{3}{2}-\epsilon
   \right)}{4^{2-2 \epsilon } \pi ^{3-2
   \epsilon}  }\int_{0}^{2\pi}\!\!\!\! d\tau_{1}\int_{\tau_{1}}^{2\pi} \!\!\!\!d\tau_{2}\int_{\tau_{2}}^{2\pi} \!\!\!\!d\tau_{3}\int_{\tau_{3}}^{2\pi} \!\!\!\!d\tau_{4}~
 \frac{ 1 }{\left[\sin
   ^2\left(\frac{\tau _2-\tau_1}{2}
  \right){}\right]^{ 1-\epsilon} \left[\sin
   ^2\left(\frac{\tau _4-\tau_3}{2}
   \right){}\right]^{1-\epsilon }},
\end{align}
and
\begin{align}
\!\!(\mathbf{II})=&\frac{\Gamma^{2}
   \left(\frac{3}{2}-\epsilon
   \right) }{4^{2-2 \epsilon } \pi ^{3-2\epsilon } } 
   \int_{0}^{2\pi}\!\!\!\! d\tau_{1}\int_{\tau_{1}}^{2\pi} \!\!\!\!d\tau_{2}\int_{\tau_{2}}^{2\pi} \!\!\!\!d\tau_{3}\int_{\tau_{3}}^{2\pi} \!\!\!\!d\tau_{4}~
   \frac{\sin\left(\frac{\tau _2-\tau_3}{2}\right) \sin
   \left(\frac{\tau _1-\tau_4}{2}
   \right) }{\left[\sin
   ^2\left(\frac{\tau _2-\tau_3}{2}
   \right){}\right]^{\frac{3}{2}-\epsilon } \left[\sin
   ^2\left(\frac{\tau _1-\tau_4}{2}
 \right){}\right]^{\frac{3}{2}-\epsilon }}=\nonumber\\
 =&\frac{\Gamma^{2}
   \left(\frac{3}{2}-\epsilon
   \right) }{4^{2-2 \epsilon } \pi ^{3-2\epsilon } } 
   \int_{0}^{2\pi}\!\!\!\! d\tau_{1}\int_{\tau_{1}}^{2\pi} \!\!\!\!d\tau_{2}\int_{\tau_{2}}^{2\pi} \!\!\!\!d\tau_{3}\int_{\tau_{3}}^{2\pi} \!\!\!\!d\tau_{4}~
   \frac{1 }{\left[\sin
   ^2\left(\frac{\tau _3-\tau_2}{2}
   \right){}\right]^{1-\epsilon } \left[\sin
   ^2\left(\frac{\tau _4-\tau_1}{2}
 \right){}\right]^{1-\epsilon }}.
\end{align}
There are different way of computing the integrals  ({\bf I}) and  ({\bf II}). The most direct is to introduce  the auxiliary function
\be
F(\tau)=-\int^{2\pi}_{\tau}\!\!\! d\tau_{1}\int^{2\pi}_{\tau_{1}}\!\!\! d\tau_{2}\int^{2\pi}_{\tau_{2}}\!\!\! d\tau_{3}~\frac{1}{[\sin^{2}\frac{\tau_{3}}{2}]^{1-\epsilon}}=-\frac{1}{2}\int^{2\pi}_{\tau}\!\!\! d\tau_{1}
\frac{(\tau_{1}-\tau)^{2}}{[\sin^{2}\frac{\tau_{1}}{2}]^{1-\epsilon}},
\ee
whose properties are discussed in detail in app. \ref{Ftau}. If we observe that
\be
F^{\prime\prime\prime}(\tau)=\frac{1}{[\sin^{2}\frac{\tau}{2}]^{1-\epsilon}},
\ee
we can easily express 
the integral  ({\bf I})  in terms of  the value of the function $F$ and its derivatives in zero. In fact, after performing a certain number of trivial integrations  and  using the basics properties given in app. \ref{Ftau}, we find
the simple expression
\be
\begin{split}
\mathrm{(\mathbf{I})}
 =&- \frac{\Gamma^{2}
   \left(\frac{3}{2}-\epsilon
   \right)}{4^{2-2 \epsilon } \pi ^{3-2
   \epsilon}  }\left(F(0) F''(0)-\frac{1}{2}
   F'(0)^2\right)=-\frac{1}{4}+O(\epsilon)
 \end{split}
\ee
The last equality can be obtained by means of  the explicit results in app.   \ref{Ftau}. The integral ($\mathbf{II}$)
can be computed along the same line and one obtains 
\begin{align}
(\mathbf{II})
 =&\frac{\Gamma^{2}
   \left(\frac{3}{2}-\epsilon
   \right) }{4^{2-2 \epsilon } \pi ^{3-2\epsilon } }\left(\frac{5}{2} \pi ^2 F''(0)^2-2
   F(0) F''(0)-\int_{0}^{2\pi}\!\!\!\! d\tau  \left(2 \pi -\tau \right)
   F^{\prime\prime}\left(\tau \right)^{2}\right)=\frac{1}{8}+O(\epsilon),
\end{align}
the last equality is again derived with the help of  app. \ref{Ftau}  where the value of each term is spelled out. The total result for the upper block is then
\be
\frac{\pi}{\kappa^2}\frac{M N}{M+N}\left( M+\frac{1}{2} N\right).
\ee
Since we are only considering diagrams which involve fermionic propagators, we find instructive  analyze the origin of this  contribution  also in terms the Ward identity \eqref{Ward3}. In particular we want to understand  the effect of the anomalous term when computing the different integrals.

For this purpose, it is more convenient to write down the r.h.s of the  Ward identity in terms of the function  
\be
G(\tau_{1}-\tau_{2})=\frac{\Gamma
   \left(\frac{1}{2}-\epsilon \right)}{2^{2-2\epsilon } \pi ^{\frac{3}{2}-\epsilon
   }}
\frac{\ \ \left[\sin
   ^2\frac{\tau _1-\tau
   _2}{2} \right]^{
   \frac{1}{2}+\epsilon}}{\sin\frac{\tau _1-\tau
   _2}{2} },
\ee
which appears in the anomalous term.   Then eq. \eqref{Ward3} can be rewritten as follows
\be
\label{Ward5}
\langle(\eta\bar\psi)_{1}(\psi\bar\eta)_{2}\rangle_{0}
=\frac{1}{\epsilon}\partial_{\tau_{1}}\partial_{\tau_{2}}G(\tau_{1}-\tau_{2})
   -\epsilon G(\tau_{1}-\tau_{2}).
\ee
We start by considering again the contribution ($\mathbf{I}$):  if we use  the above identity   it can be rewritten as follows
\begin{align}
\label{rtuy}
&\int_{0}^{2\pi}\!\!\!\! \!d\tau_{1}\int_{\tau_{1}}^{2\pi} \!\!\!\!\!d\tau_{2}\int_{\tau_{2}}^{2\pi} \!\!\!\!\!d\tau_{3}\int_{\tau_{3}}^{2\pi} \!\!\!\!\!d\tau_{4}~
\langle (\eta\bar\psi)_{1}(\psi\bar\eta)_{2}\rangle_{0} \langle(\eta\bar\psi)_{3}(\psi\bar\eta)_{4}\rangle_{0}=
\nn\\
   =&\int_{0}^{2\pi}\!\!\!\! \!d\tau_{1}\int_{\tau_{1}}^{2\pi} \!\!\!\!\!d\tau_{2}\int_{\tau_{2}}^{2\pi} \!\!\!\!\!d\tau_{3}\int_{\tau_{3}}^{2\pi} \!\!\!\!\!d\tau_{4}~\left[\frac{1}{\epsilon^{2}} \partial_{\tau_{1}}\partial_{\tau_{2}} G(\tau_{1}-\tau_{2}) \partial_{\tau_{3}}\partial_{\tau_{4}}G(\tau_{3}-\tau_{4})-\right.\\
   &- G(\tau_{1}-\tau_{2}) \partial_{\tau_{3}}\partial_{\tau_{4}}G(\tau_{3}-\tau_{4})-
\partial_{\tau_{1}}\partial_{\tau_{2}} G(\tau_{1}-\tau_{2}) G(\tau_{3}-\tau_{4})+\epsilon^{2}G(\tau_{1}-\tau_{2})G(\tau_{3}-\tau_{4})\biggr].\nn
\end{align}
We separate eq. \eqref{rtuy} into three different contributions, according to the powers of $\epsilon$  which appear explictly in the integral. First we consider the $\epsilon^{2}$ part: it is the product of two anomalous terms.
In the limit $\epsilon\to 0$ it  identically vanishes  since $G$ becomes a constant.  Next we consider    the $1/\epsilon^{2}$ contribution in eq. \eqref{rtuy}, which would  be the only one present if the  Ward identity were valid in the {\it classical} form \eqref{Ward2}. Using the explicit total  derivatives acting on $G$ we can easily perform three of the four integrations and we get
\begin{align}
&\frac{1}{\epsilon^{2}} \int_{0}^{2\pi}\!\!\!\! \!d\tau\left((2 \pi -\tau ) G'(0)-G(\tau
   )+G(2 \pi )\right)
   \left(G'(0)-G'(\tau )\right)=\nn\\
   =& \frac{1}{\epsilon^{2}} \int_{0}^{2\pi}\!\!\!\! \!d\tau G(\tau
   )
   G'(\tau )=\frac{1}{2\epsilon^{2}} [(G(2\pi))^{2}-(G(0))^{2}]=0.
\end{align}
In  dimensional regularization we can always work in the region of the 
$\epsilon-$plane where
 $G(0)=G(2\pi)=G^{\prime}(0)=G^{\prime}(2\pi)=0$. Finally we consider the $\epsilon^{0}$-terms. They arise when the classical term in the Ward identity is multiplied by the anomalous one. Since they contain two total 
 derivative we can eliminate two integrations and we get
\begin{align}
&-2 \int_{0}^{2\pi}\!\!\!\! \!d\tau_{1}\int_{\tau_{1}}^{2\pi} \!\!\!\!\!d\tau_{2}G\left(\tau _1-\tau _2\right)
   G\left(\tau _2\right)=(\epsilon\to 0)=-\frac{2}{16\pi^{2}} \int_{0}^{2\pi}\!\!\!\! \!d\tau_{1}\int_{\tau_{1}}^{2\pi} d\tau_{2}~ 1=-\frac{1}{4}.
\end{align}
Summarizing the non-vanishing result completely originates from the anomalous term: in its absence it would  be identically zero!  The anomalous term is no longer a spectator in our computation.

The other contribution  can be analyzed in a similar way. In fact its expansion in terms of $G-$function is
\begin{align}
\label{3.31}
&\int_{0}^{2\pi}\!\!\!\! \!d\tau_{1}\int_{\tau_{1}}^{2\pi} \!\!\!\!\!d\tau_{2}\int_{\tau_{2}}^{2\pi} \!\!\!\!\!d\tau_{3}\int_{\tau_{3}}^{2\pi} \!\!\!\!\!d\tau_{4}~
\langle (\eta\bar\psi)_{1}(\psi\bar\eta)_{4}\rangle_{0} \langle (\eta\bar\psi)_{3}(\psi\bar\eta)_{2}\rangle_{0}=\nn\\
   =&\int_{0}^{2\pi}\!\!\!\! \!d\tau_{1}\int_{\tau_{1}}^{2\pi} \!\!\!\!\!d\tau_{2}\int_{\tau_{2}}^{2\pi} \!\!\!\!\!d\tau_{3}\int_{\tau_{3}}^{2\pi} \!\!\!\!\!d\tau_{4}\left[\frac{1}{\epsilon^{2}} \partial_{\tau_{1}}\partial_{\tau_{4}} G(\tau_{1}-\tau_{4}) \partial_{\tau_{3}}\partial_{\tau_{2}}G(\tau_{3}-\tau_{2})-\right.\\
   &-G(\tau_{1}-\tau_{4}) \partial_{\tau_{3}}\partial_{\tau_{2}}G(\tau_{3}-\tau_{2})-
\partial_{\tau_{1}}\partial_{\tau_{4}} G(\tau_{1}-\tau_{4}) G(\tau_{3}-\tau_{2})+\epsilon^{2}G(\tau_{1}-\tau_{4})G(\tau_{3}-\tau_{2})\biggr].\nn
\end{align}
Again the  terms proportional to $\epsilon^{2}$ can be neglected in 
the limit $\epsilon\to 0$ and in the contribution of order $0$ in $\epsilon$ 
we can eliminate two of the four integration. For the latter we find in fact
\begin{align}
 &-\int_{0}^{2\pi}\!\!\!\! \!d\tau_{1}\int_{\tau_{1}}^{2\pi} \!\!\!\!\!d\tau_{4} G\left(\tau _4-\tau _1\right)
   \left(2 G\left(\tau _4-\tau
   _1\right)-G\left(\tau
   _4\right)-G\left(\tau
   _1\right)\right)=\nn\\
   =&(\epsilon\to 0)=-\frac{1}{16\pi^{2}} \int_{0}^{2\pi}\!\!\!\! \!d\tau_{1}\int_{\tau_{1}}^{2\pi} d\tau_{4} [2-1-1]=0.
      \end{align}
Therefore all the contributions coming from the anomalous term in the Ward identity \eqref{Ward3} vanish for the integral ($\mathbf{II}$)  in \eqref{DEX}. The actual value of ($\mathbf{II}$) is only determined by the first term in \eqref{3.31}. We find in fact
\begin{align}
&\frac{1}{\epsilon^{2}}\int_{0}^{2\pi}\!\!\!\! \!d\tau_{1}\int_{\tau_{1}}^{2\pi} \!\!\!\!\!d\tau_{2}\int_{\tau_{2}}^{2\pi} \!\!\!\!\!d\tau_{3}\int_{\tau_{3}}^{2\pi} \!\!\!\!\!d\tau_{4}
\partial_{\tau_{1}}\partial_{\tau_{4}}G(\tau_{1}-\tau_{4}) \partial_{\tau_{3}}\partial_{\tau_{2}}G(\tau_{3}-\tau_{2})=\nn\\
&=\frac{1}{\epsilon^{2}}\int_{0}^{2\pi}\!\!\!\! \!d\tau_{3}\int_{0}^{\tau_{3}} \!\!\!\!\!d\tau_{2} G(\tau_{3}-\tau_{2})G^{\prime\prime}(\tau_{3}-\tau_{2})=
\frac{\pi}{\epsilon^{2}}\int_{0}^{2\pi}\!\!\!\! \!d\tau G(\tau)G^{\prime\prime}(\tau)=\frac{1}{8}+O(\epsilon).
\end{align}

\subsection{ Vertex contribution}
The last set of  fermionic diagrams that we have to consider in our perturbative evaluation  originates
when  we take into account the familiar gauge-interaction between  fermion and  
\label{Vertdiag}
\begin{wrapfigure}[9]{l}{60mm}
\centering{
    \includegraphics[width=.2\textwidth, height=.2\textwidth]{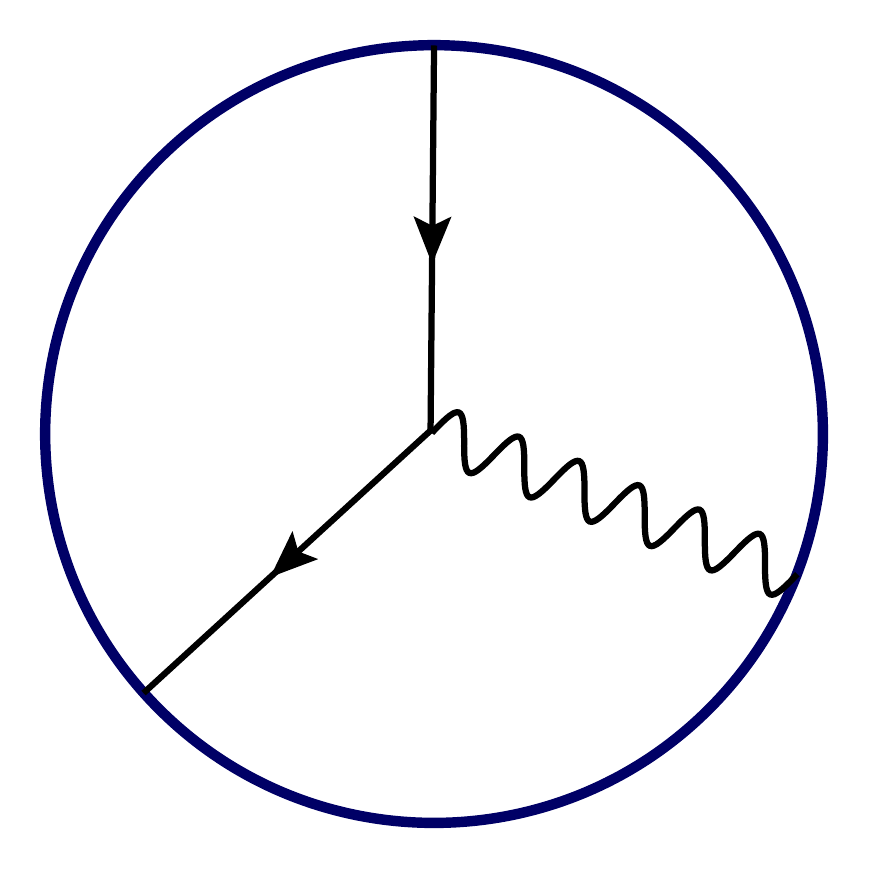}
   }
\caption{\label{Vertexa} Fermionic vertex diagram.}
\end{wrapfigure}
 vector fields. 
With respect to the cases analyzed in the previous subsections, this contribution involves an additional complications:  the integration over the position of a  gauge-spinor-spinor vertex (see fig. \ref{Vertexa}).  These diagrams  arise when expanding
the three cubic monomial   in the second line of  \eqref{expaloop}, {\it i.e.} 
\begin{align}
&-\left(\frac{2\pi}{\kappa}\right)\frac{ i}{M+N} \oint \!d\tau_{\mbox{\tiny $\displaystyle 3\!\!>\!\! 2\!\!>\!\!1$}}~~\mathrm{Tr}
\Bigl[\langle (\eta\bar{\psi})_1(\psi\bar{\eta})_2{\cal A}_3 \rangle
+\nn\\
&+\langle (\eta\bar{\psi})_1\hat{\cal A}_2 (\psi\bar{\eta})_3\rangle+\langle{\cal A}_1(\eta\bar{\psi})_2(\psi\bar{\eta})_3\rangle\Bigr],
\end{align}
where the expectation values are obviously taken in the interacting theory.  Inserting 
the relevant interaction Lagrangian in the $\langle\cdots\rangle$ and taking all the  Wick contractions, the contribution of these three monomial  can  be rearranged as follows
\begin{align}
\label{calSp}
\mathcal{S}=& \left(\frac{2\pi}{\kappa}\right)^{2}\frac{1}{M+N}\int_{0}^{2\pi}\!\!\!d\tau_{\mbox{\tiny $\displaystyle 3\!\!>\!\! 2\!\!>\!\!1$}}~\left[N^{2} M(\eta_{1L}\gamma_{\nu}\gamma^{\mu}\gamma_{\lambda}\bar\eta_{2}^{L}) \epsilon_{\mu\rho\sigma}\dot{x}_{3}^{\rho}~\Gamma^{\nu\lambda\sigma}+\right.\nn\\
&+\left.N^{2} M(\eta_{2L}\gamma_{\lambda}\gamma^{\mu}\gamma_{\nu}\bar\eta^{L}_{3})
\epsilon_{\mu\rho\sigma}\dot{x}_{1}^{\rho}~\Gamma^{\sigma\lambda\nu}+NM^{2}(\eta_{1L}
\gamma_{\lambda}\gamma^{\mu}\gamma_{\nu}
 \bar\eta_{3}^{\ L})\epsilon_{\mu\rho\sigma}\dot{x}_{2}^{\rho}
~ \Gamma^{\lambda\sigma\nu}\right],
\end{align}
where we have also performed the trace over all the gauge indices and used that $|\dot x_{i}|=1$ for the circle. The function  $\Gamma^{\lambda\mu\nu}$
is a short-hand notation which hides the three-point function defined by the integral
\be
\label{threepoint}
\begin{split}
\Gamma^{\lambda\mu\nu}(x_{1},x_{2}, x_{2})=&\left(\frac{\Gamma(\frac{1}{2}-\epsilon)}{4\pi^{3/2-\epsilon}}\right)^{3}\partial_{x_{1}^{\lambda}}\partial_{x_{2}^{\mu}}\partial_{x_{3}^{\nu}}
\int 
\frac{d^{3-2\epsilon}w}{(x_{1w}^{2})^{1/2-\epsilon}(x_{2w}^{2})^{1/2-\epsilon}(x_{3w}^{2})^{1/2-\epsilon}}.
\end{split}
\ee
In this case it is convenient to sum upper and lower block, before performing the integration over the contour because  cancellations between  objects appearing in different blocks occur. We can rewrite the total contribution as
\begin{align}
\label{calS}
\mathcal{S}=& \left(\frac{2\pi}{\kappa}\right)^{2}{MN}\int_{0}^{2\pi}\!\!\!d\tau_{\mbox{\tiny $\displaystyle 3\!\!>\!\! 2\!\!>\!\!1$}}~\left[\underset{(A_{12;3)}}{(\eta_{1L}\gamma_{\nu}\gamma^{\mu}\gamma_{\lambda}\bar\eta_{2}^{L}) \epsilon_{\mu\rho\sigma}\dot{x}_{3}^{\rho}~\Gamma^{\nu\lambda\sigma}}+\right.\nn\\
&+\left.\underset{(A_{23;1)}}{(\eta_{2L}\gamma_{\lambda}\gamma^{\mu}\gamma_{\nu}\bar\eta^{L}_{3})
\epsilon_{\mu\rho\sigma}\dot{x}_{1}^{\rho}~\Gamma^{\sigma\lambda\nu}}+\underset{A_{(13;2)}}{(\eta_{1L}
\gamma_{\lambda}\gamma^{\mu}\gamma_{\nu}
 \bar\eta_{3}^{\ L})\epsilon_{\mu\rho\sigma}\dot{x}_{2}^{\rho}
~ \Gamma^{\lambda\sigma\nu}}\right].
\end{align}
The next step is to expand the spinor bilinears in \eqref{calS} in terms of the circuit tangent vectors $\dot{x}_i$ and of the scalar spinor contractions  $\eta_i \bar\eta_j$. For instance, the first term  in \eqref{calS} can be rewritten  as follows
\be
\label{V1}
\begin{split}
\!\!\!\!A_{12;3}&=(\eta_{1L}\gamma_{\nu}\gamma^{\mu}\gamma_{\lambda}\bar\eta_{2}^{L}) \epsilon_{\mu\rho\sigma}\dot{x}_{3}^{\rho}~\Gamma^{\nu\lambda\sigma}=
(\eta_{1}\gamma_{\nu}\gamma^{\mu}\gamma_{\lambda}\bar\eta_{2}) \epsilon_{\mu\rho\sigma}\dot{x}_{3}^{\rho}~\Gamma^{\nu\lambda\sigma}=\\
&=[\delta_\nu^\mu\eta_{1}\gamma_{\lambda}\bar\eta_{2}+\delta_\lambda^\mu\eta_{1}\gamma_{\nu}\bar\eta_{2}-\delta_{\lambda\nu}\eta_{1}\gamma^{\mu}\bar\eta_{2}- i
\eta_{1}
\bar\eta_{2}\epsilon_{\  ~ \nu\lambda}^{\mu} ]\epsilon_{\mu\rho\sigma}\dot{x}_{3}^{\rho}~\Gamma^{\nu\lambda\sigma}.
\end{split}
\ee
In the second equality we have dropped  the dependence on the R-symmetry indices since their contraction simply yields a factor $1$. From now on $\eta$ and $\bar \eta$
will just describe the spinor part of eq. \eqref{coup}. The vector defined  by the fermionic bilinear $\eta_{1}\gamma_{\nu}\bar\eta_{2}$ appearing in \eqref{V1} can be
expressed  in terms of the  tangent vectors to the circuit  as illustrated in \cite{Griguolo:2012iq}. For the circle one can write
\be 
\label{V2}
\begin{split}
(\eta_{1}\gamma^{\mu}\bar\eta_{2})
=&-\frac{2}{(\eta_{2}\bar\eta_{1})}\left[
\dot{x_{1}}^{\mu}+  \dot{x_{2}}^{\mu}-i \dot{x_{1}}^{\lambda}
\dot{x_{2}}^{\nu}\epsilon_{\lambda\nu}^{\ \  \ \mu}
\right].
\end{split}
\ee
Since we are dealing with a planar circuit, lying on the plane $x_{3}=0$,   only the  last term in \eqref{V2} yields   non-vanishing results when inserted into \eqref{V1}.  After some long but trivial  tensorial manipulation, using crucially that the contour is a circle, we can rearrange this contribution in a nice way
\be
\begin{split}
A_{12;3}&=i  \left[\frac{2  }{(\eta_{2}\bar\eta_{1})} (\delta_\nu^\mu \epsilon_{\alpha\beta \lambda}\dot{x_{1}}^{\alpha}
\dot{x_{2}}^{\beta}+\delta_\lambda^\mu \epsilon_{\alpha\beta \nu}\dot{x_{1}}^{\alpha}
\dot{x_{2}}^{\beta}-\delta_{\lambda\nu} \epsilon_{\alpha\beta}^{\ \ \ \mu}\dot{x_{1}}^{\alpha}
\dot{x_{2}}^{\beta})-\right.\\  
&-(\eta_{1}
\bar\eta_{2})\epsilon_{\  ~ \nu\lambda}^{\mu} \biggr]\epsilon_{\mu\rho\sigma}\dot{x}_{3}^{\rho}~\Gamma^{\nu\lambda\sigma}
=
 i
 \biggl[(\eta_{1}\bar\eta_{2}) \dot{x}_{3\lambda}(\Gamma^{\tau\lambda\tau}-\Gamma^{\lambda\tau\tau})+\\
&+\frac{ r_{12} }{(\eta_{2}\bar\eta_{1})}[x_{23}^{2}+x_{13}^{2}] h_{\mu} h_{\nu} \mathbf{V}^{\mu\nu}
+\frac{2r_{12}}{(\eta_{2}\bar\eta_{1})} x_{3\nu}\Gamma^{\tau\tau\nu}\Biggr].
\end{split}
\ee
Let us notice that we have introduced a non-covariant notation where $h^{\mu}$ stands for  the versor $\delta^{\mu}_{3}$, while $r_{ij}=\epsilon_{\lambda\mu\nu} h^{\lambda}\dot{x_{i}}^{\mu} \dot{x_{j}}^{\nu}$. The explicit appearence in third contraction of the vector $x^\mu$ is peculiar of the fact that we are dealing with a circle. The three-point function  \eqref{threepoint} appears in two different ways in the above expression: (a) its contracted forms $\Gamma^{\nu\tau\tau},~\Gamma^{\tau\nu\tau}$  and $\Gamma^{\tau\tau\nu}$; (b) the tensor integral $\mathbf{V}^{\mu\nu}$
\be
\label{Vbmn}
\begin{split}
\mathbf{V}^{\mu\nu}
\equiv&-\left(\frac{\Gamma(\frac{3}{2}-\epsilon)}{2\pi^{3/2-\epsilon}}\right)^{3}
\int d^{3-2\epsilon}w
\frac{w^{\mu}w^{\nu}}{(x_{1w}^{2})^{3/2-\epsilon}(x_{2w}^{2})^{3/2-\epsilon}(x_{3w}^{2})^{3/2-\epsilon}}.
\end{split}
\ee
As discussed in \cite{Griguolo:2012iq} the contracted three-point functions greatly simplify and reduce to the derivatives of product of scalar propagator [see \cite{Griguolo:2012iq} for the details]: we expect that the related integrals will be similar to the ones appearing in the double exchange computation. The only point where the use of actual 
three-point function  in $D=3-2\epsilon$ seems instead to be unavoidable  is when we consider the contribution of $\mathbf{V}^{\mu\nu}$.

Below, we shall show that this complication can be avoided following the approach discussed in \cite{Bassetto:2008yf}. The idea is to regularize  the behavior 
of  the integral of $h_{\mu} h_{\nu} \mathbf{V}^{\mu\nu}$ at coincident points by an appropriate subtraction which removes the unwelcome singular behavior. A useful subtraction must respect two criteria: (1) it can be expressed in terms of two point functions;  (2) it eliminates the singular behavior for $\epsilon\ne 0$ and not just in the limit $\epsilon\to 0$.  This second  property is particularly important, otherwise  one can lose  finite terms arising from the product of divergent quantities with evanescent terms (see also \cite{Bianchi:2013pva} to appreciate the importance of this kind of contributions in a different but related context).

In appendix \ref{AppA1} it is shown, by studying carefully the behavior of the integral  at coincident points,  that  the combination
\be
r_{12}[(x_{12}^2+x_{13}^2)h_{\mu} h_{\nu} \mathbf{V}^{\mu\nu}-2 (1-2\epsilon) x_{1\mu}\Gamma^{\mu\rho\rho}]
\ee
is completely regular when either   $x_{12}^2$ or $x^{2}_{13}$ or $x_{23}^2$ approaches to zero. We remark the appearence of an explicit dependence, in our subtraction term, from the $\epsilon$ parameter, exactly of evanescent type: it is essential to obtain a complete regular result at coincident points. We can finally rearrange the expression for $A_{12;3}$  as follows 
\be
\label{(A)}
\begin{split}
A_{12;3}&
=
 i
 \biggl[(\eta_{1}\bar\eta_{2}) \dot{x}_{3\lambda}(\Gamma^{\tau\lambda\tau}-\Gamma^{\lambda\tau\tau})+
 \\
&+\frac{ r_{12} }{(\eta_{2}\bar\eta_{1})}[(x_{23}^{2}+x_{13}^{2}) h_{\mu} h_{\nu} \mathbf{V}^{\mu\nu}-
2(1-2\epsilon) x_{3\nu}\Gamma^{\tau\tau\nu}]
+\frac{4 (1-\epsilon)r_{12}}{(\eta_{2}\bar\eta_{1})} x_{3\nu}\Gamma^{\tau\tau\nu}\Biggr],
\end{split}
\ee
in order to single out the regular combination. The singular behavior, once present 
in the three point function $h_{\mu} h_{\nu} \mathbf{V}^{\mu\nu}$, has been  now shifted in the last term of \eqref{(A)}, which  is expressible just in terms of derivative of products of scalar
propagators.  With this simple trick the only  integrals that we have to perform in $D=3-2\epsilon$  are those involving  the contracted three point functions.

We shall obviously perform the same procedure on the other two contributions  appearing in \eqref{calS}. We get
\begin{subequations}
\begin{align}
A_{23;1}&
=
 i
 \biggl[(\eta_{2}\bar\eta_{3}) \dot{x}_{1\nu}(\Gamma^{\tau\tau\nu}-\Gamma^{\tau\nu\tau})+\nn\\
&+\frac{ r_{23} }{(\eta_{3}\bar\eta_{2})}[(x_{13}^{2}+ x_{12}^{2}) h_{\mu} h_{\nu} \mathbf{V}^{\mu\nu}-2(1-2\epsilon) x_{1\nu}\Gamma^{\nu\tau\tau}]
+\frac{4(1-\epsilon) r_{23}}{(\eta_{3}\bar\eta_{2})}x_{1\nu}\Gamma^{\nu\tau\tau}\Biggr],
\\
A_{13;2}&
=
 i
 \biggl[(\eta_{1}\bar\eta_{3})\dot{x}_{2\nu}(\Gamma^{\tau\tau\nu}-\Gamma^{\nu\tau\tau})+\nn\\
&+\frac{ r_{13} }{(\eta_{3}\bar\eta_{1})}[(x_{12}^{2}+ x_{23}^{2}) h_{\mu} h_{\nu} \mathbf{V}^{\mu\nu}-
2(1-2\epsilon) x_{2\nu}\Gamma^{\tau\nu\tau}]
+\frac{4(1-\epsilon)r_{13}}{(\eta_{3}\bar\eta_{1})} x_{2\nu}\Gamma^{\tau\nu\tau}\Biggr].
\end{align}
\end{subequations}
We are ready to compute the integrals and we begin  from those we can directly evaluate for $\epsilon=0$.  We  consider in fact the combination
\begin{align}
-i \left(\frac{2\pi}{\kappa}\right)^{2}N M\int_{0}^{2\pi} \!\!\! \!\!\! d\tau_{1}\int_{\tau_{1}}^{2\pi}\!\!\! \!\!\! d\tau_{2}
&\int_{\tau_{2}}^{2\pi}\!\!\! \!\!\! d\tau_{3}\biggl[\frac{ r_{12} }{(\eta_{2}\bar\eta_{1})}[(x_{23}^{2}+x_{13}^{2}) h_{\mu} h_{\nu} \mathbf{V}^{\mu\nu}-
2(1-2\epsilon) x_{3\nu}\Gamma^{\tau\tau\nu}]+\nn\\
&+\frac{ r_{23} }{(\eta_{3}\bar\eta_{2})}[(x_{13}^{2}+ x_{12}^{2}) h_{\mu} h_{\nu} \mathbf{V}^{\mu\nu}-2(1-2\epsilon) x_{1\nu}\Gamma^{\nu\tau\tau}]
+\nn\\
&+\frac{ r_{13} }{(\eta_{3}\bar\eta_{1})}[(x_{12}^{2}+ x_{23}^{2}) h_{\mu} h_{\nu} \mathbf{V}^{\mu\nu}-
2(1-2\epsilon) x_{2\nu}\Gamma^{\tau\nu\tau}]\biggr],
\end{align}
which for $\epsilon=0$ reduces to this simple integral
\be
\label{V1a}
-\frac{M N}{32\kappa^{2}}\int_{0}^{2\pi} \!\!\!\!\!d\tau_{1}\int_{\tau_{1}}^{2\pi} \!\!\!\!\! d\tau_{2}
\int_{\tau_{2}}^{2\pi} \!\!\!\!\! d\tau_{3}
\frac{ \cos
  \frac{\tau
   _1-\tau _2}{2}- \cos
   \frac{\tau
   _1-\tau _3}{2} + \cos
   \frac{\tau
   _2-\tau _3}{2}+\frac{3}{2}
}{   \sin\frac{\tau_1-\tau _3}{4} \cos
   \frac{\tau_1-\tau _2}{4} \cos
   \frac{\tau
   _2-\tau _3}{4}}=-\frac{\pi ^2 M N }{2
   \kappa ^2}(1-4\log 2).
\ee
Next we consider the two type of contribution containing contracted three point functions: we have
\begin{align}
\label{Cab11}
I_{1}=i\left(\frac{2\pi}{\kappa}\right)^{2}\!\!\!N M\!\!\int_{0}^{2\pi} \!\!\! \!\!\! d\tau_{1}\!\int_{\tau_{1}}^{2\pi}\!\!\! \!\!\! d\tau_{2}\!
\int_{\tau_{2}}^{2\pi}\!\!\! \!\!\! d\tau_{3}&\left[(\eta_{1}\bar\eta_{2}) \dot{x}_{3\lambda}(\Gamma^{\tau\lambda\tau}-\Gamma^{\lambda\tau\tau})+(\eta_{2}\bar\eta_{3}) \dot{x}_{1\nu}(\Gamma^{\tau\tau\nu}-\Gamma^{\tau\nu\tau})+\right.\nn\\
&+\left.(\eta_{1}\bar\eta_{3})\dot{x}_{2\nu}(\Gamma^{\tau\tau\nu}-\Gamma^{\nu\tau\tau})\right].
\end{align}
and
\be\label{Cab12}
\mbox{\small$\displaystyle
I_{2}=4i(1-\epsilon) \left(\frac{2\pi}{\kappa}\right)^{2}\!\!\!N M\!\!\int_{0}^{2\pi} \!\!\! \!\!\! d\tau_{1}\!\int_{\tau_{1}}^{2\pi}\!\!\! \!\!\! d\tau_{2}\!
\int_{\tau_{2}}^{2\pi}\!\!\! \!\!\! d\tau_{3}\!\left[\frac{r_{12}}{(\eta_{2}\bar\eta_{1})} x_{3\nu}\Gamma^{\tau\tau\nu}+\frac{ r_{23}}{(\eta_{3}\bar\eta_{2})}x_{1\nu}\Gamma^{\nu\tau\tau}+\frac{r_{13}}{(\eta_{3}\bar\eta_{1})} x_{2\nu}\Gamma^{\tau\nu\tau}\right].$}
\ee
These integrals are computed quite easily, once the full simmetry of the expressions is exploited and the explicit reduction in terms of two-point function has been considered. In particular we have used the definition of the function $\Phi_{i,jk}$ (that was given in  app.  B of \cite{Griguolo:2012iq} and whose basic properties are recollected in Appendix (\ref{ABJ})) and a peculiar property of the circle: the invariance under translation of the function $\Phi_{i,jk}$ with respect to the  contour parameters $\tau_i$. We also have taken advantage of some total derivative terms. The details of the computations are deferred to Appendix (\ref{AppD}). We finally get:
\begin{align}
I_{1}=\label{V3}
=&-\frac{M N 4^{2 \epsilon -1} \pi ^{2 \epsilon +\frac{1}{2}} \Gamma \left(\frac{1}{2}-\epsilon \right)^2 \Gamma \left(2 \epsilon +\frac{1}{2}\right)}{\kappa ^2 \epsilon ^2 \Gamma (2 \epsilon )}=\nn\\
=&-\frac{\pi ^{2+2\epsilon} M N }{2 \kappa ^2  }\left(\frac{1}{\epsilon} +2 \gamma   +  4\log 2+O(\epsilon)\right).
\end{align}
and
\begin{align}
\label{V22a}
I_{2}=&-\frac{M N 4^{2 \epsilon -1} \pi ^{2
   \epsilon } \Gamma
   \left(\frac{1}{2}-\epsilon
   \right)^2 \left(\pi  \epsilon 
   \Gamma (2 \epsilon )-\sqrt{\pi }
   \Gamma \left(2 \epsilon
   +\frac{1}{2}\right)\right)}{\kappa
   ^2 \epsilon ^2 \Gamma (2 \epsilon
   -2)}=\\
   =&-\frac{\pi ^{2+2\epsilon} M N }{2 \kappa ^2 
   }\left(\!-\frac{1}{\epsilon} +3-2 \gamma+O(\epsilon)\!\right)\!.\nn
\end{align}

Finally if we sum the three contributions \eqref{V1a},  \eqref{V3} and \eqref{V22a} we obtain the following  result for the whole vertex
\begin{align}
&-\frac{\pi ^2 M N }{2
   \kappa ^2}\left[1\!-\!4\log 2\!+\!\pi^{2\epsilon}\! \left(\!-\frac{1}{\epsilon} \!+\!3\!-\!2 \gamma\!+\!O(\epsilon)\!\right)\!+\!\pi ^{2\epsilon} \left(\frac{1}{\epsilon}\! +\!2 \gamma \! + \! 4\log 2\!+\!O(\epsilon)\right)\right]_{\epsilon\to 0}=\nn\\
   =&-
   \frac{2\pi ^2 M N }{
   \kappa ^2}.
\end{align}
\section{Comparison with the localization result}
We have to sum now the contribution of the lower block and to collect all the results: we get the following prediction at two-loop from the perturbative computation
\be
\begin{split}
\langle{\cal W}_F\rangle^{f=0}=&1+\underset{\rm Bubbles}{\frac{\pi^{2}}{\kappa} M N} -\underset{\rm Gauge ~Vertex}{\frac{\pi^{2}}{6\kappa}(M^{2}+N^{2}-MN)}+\underset{\rm Fermionic~ DE}{\frac{3}{2}  \frac{\pi^{2}}{\kappa} MN}-\underset{\rm Fermion~ Vertex}{2 \frac{\pi^{2}}{\kappa} MN}\\
=&1-\frac{\pi^{2}}{6\kappa^{2}} (M^{2}-4 MN +N^{2}) ,
\end{split}
\ee
which is exactly  the outcome of the matrix model analysis \eqref{fram1/2}, when the framing phase is stripped off \cite{Drukker:2010nc}. We see clearly how the fermionic interactions play a decisive role in recovering the localization result. It is also manifest the violation, at framing zero, of the cohomological equivalence in conventional perturbation theory, where evanescent terms enter crucially into the game.
\section{Conclusions}
We have computed at two-loop the quantum expectation value of the 1/2 BPS Wilson loop using DRED regularization. We have implicitly worked at framing zero and we have found full consistency with the localization results, that are produced at framing one. In order to compare the two expressions, one has to single out the framing phase into the matrix model outcome: this has been done in \cite{Drukker:2010nc} and their framing zero formula precisely coincides with our perturbative answer. On the other hand, at framing zero the cohomological equivalence between 1/2 BPS and 1/6 BPS Wilson loops is expected to be violated and fermions should contribute actively to the quantum averages. We have observed exactly a non-vanishing result for the sum of fermionic double exachange diagrams and gauge-fermion-fermion vertex: we have also seen how evanescent terms, in DRED regularization, play an important role in obtaining the matrix model expression. In the case of the fermionic double exchanges we have explicitly shown that they appear as anomalous contributions to a Ward identity controlling the cohomological equivalence. It would be nice to understand this feature also in the vertex diagram. Concerning the actual vertex computation, we have followed the strategy presented in \cite{Bassetto:2008yf}, performing an useful subtraction that allowed us to reduce the problem to simple propagator-like terms plus a truly vertex-like integral. This last piece is perfectly finite and can be computed directly in $D=3$, simplifying enormously the calculation. In \cite{Bassetto:2008yf} we have been able to obtain the subtracted integral for a vast class of supersymmetric loops in four dimensional ${\cal N}=4$ super Yang-Mills, namely the DGRT loops on $S^2$ \cite{Drukker:2007qr}. It would be interesting to understand if such technique could be applied to more general situation in three-dimensions. In \cite{Cardinali:2012ru} we have in fact introduced two new families of Wilson loop operators in ${\cal N} = 6$ supersymmetric Chern–Simons theory. The first one is defined for an arbitrary contour on the three-dimensional space and it resembles the Zarembo construction \cite{Zarembo:2002an} in four-dimensional ${\cal N} = 4$ super Yang-Mills. The second one involves arbitrary curves on the two dimensional sphere. In both cases one can add certain scalar and fermionic couplings to the Wilson loop so it preserves at least two supercharges. The study at quantum level of these families, using both perturbation theory and localization, will be the subject of future investigations   

\section*{Acknowledgements}
This work was supported in part by the MIUR-PRIN contract 2009-KHZKRX. We warmly thank Marco Bianchi and Silvia Penati for useful discussion and sharing their results with us before publication. We are also grateful to Nadav Drukker and Diego Trancanelli for drawing our attention to this problem.
 \newpage
\appendix
\addcontentsline{toc}{section}{Appendices}
\begin{flushleft}
\bf \large Appendices
\end{flushleft}

\section{Conventions and outlook}
\label{ABJ}
First of all we shall  summarize some basic  features of the ABJ(M)  theories  in Euclidean space-time. The gauge sector consists of  two  gauge fields $(A_\mu)_{i}^{\ j}$ and
$(\hat{A}_\mu)_{\hat{i}}^{\ \hat{j}}$ belonging respectively  to the adjoint of $U(N)$ and $U(M)$. The
matter sector  instead contains   the complex fields $(C_I)_{i}^{\ \hat{i}}$ and $({\bar C}^{I})_{\hat{i}}^{\  i}$ as well as the fermions $(\psi_I)_{\hat{i} }^{ \ i}$ and $({\bar\psi}^{I})_{i }^{\ \hat{i} }$. The fields $(C,\bar \psi)$ transform in  the $({\bf N},{\bf \bar M})$  of the gauge group $U(N)\times U(M)$ while the couple $(\bar C, \psi)$ lives in the $({\bf \bar N},{\bf M})$. The additional  capital index 
$I=1,2,3,4$  belongs to the $R-$symmetry group $SU(4)$.  In order to quantize the theory at the perturbative level, we have introduced  the covariant gauge fixing function 
$\partial_\mu A^\mu$ for both gauge fields and two sets of ghosts $(\bar c,c)$ and
$(\bar{\hat c},\hat c)$. We work therefore with the following Euclidean space action 
(see \cite{Chen:1992ee,Aharony:2008ug,Benna:2008zy})
\begin{align}
\label{Lagra}
S_\text{CS} & = -i\frac{ k}{4\pi}\,\int d^3x\, \varepsilon^{\mu\nu\rho} \,\Bigl [\,
\mathrm{Tr} (A_\mu\partial_\nu A_\rho+\frac{2}{3}\,A_\mu A_\nu A_\rho)- 
\mathrm{Tr} (\hat{A}_\mu\partial_\nu \hat{A}_\rho+\frac{2}{3}\, \hat{A}_\mu \hat{A}_\nu \hat{A}_\rho)\, \Bigr ] \nonumber\\
S_\text{gf} & = \frac{k}{4\pi}\, \int d^3 x\, \Bigl [\,\frac{1}{\xi}\, \mathrm{Tr}(\partial_\mu A^\mu)^2
+\mathrm{Tr}(\partial_\mu \bar c\, D_\mu c) - \frac{1}{\xi}\, \mathrm{Tr}(\partial_\mu \hat{A}^\mu)^2
+\mathrm{Tr}(\partial_\mu \bar{ \hat c}\, D_\mu \hat c) \, \Bigr ]\nonumber\\
S_\text{Matter} & = \int d^3 x\, \Bigl [\, \mathrm{Tr}(D_\mu\, C_I\, D^\mu {\bar C}^{I}) + i\, \mathrm{Tr}(\bar\psi^I
\, \slsh{D}\, \psi_I)\, \Bigr ] + S_\text{int}.
\end{align}
Here $S_\text{int}$ consists of the sextic scalar potential and $\psi^2 C^2$ Yukawa type potentials
spelled out in \cite{Aharony:2008ug}. The matter covariant derivatives are defined as
\be
\begin{aligned}
D_\mu C_I &= \partial_\mu C_I +i (A_\mu\, C_I - C_I\, \hat{A}_\mu) \\ 
D_\mu {\bar C}^{I} &= \partial_\mu {\bar C}^{I} -i ({\bar C}^{I}\,A_\mu -  \hat{A}_\mu\, {\bar C}_{I})  \\ 
D_\mu \psi_I &= \partial_\mu \psi_I -i (\hat{A}_\mu\, \psi_I - \psi_I\, A_\mu)  \\ 
D_\mu {\bar\psi}^{I} &= \partial_\mu {\bar \psi}^{I} +i ({\bar \psi}^{I}\,\hat{A}_\mu -  A_\mu\, 
{\bar \psi}^{I})\, . 
\end{aligned}
\label{cov-deriv}
\ee
\paragraph{\sc Feynman rules:}  Next we shall briefly summarize the 
Euclidean Feynman rules relevant for our computation and some general conventions.   We use the 
position-space propagators, which  are obtained  from those in momentum space (see {\it e.g.} \cite{Drukker:2008zx}) by means of  the following  master integral
\be
\int \frac{d^{3-2\epsilon} p}{(2\pi)^{3-2\epsilon}} \frac{e^{i p\cdot x}}{(p^{2})^{s}}=\frac{\Gamma\left(\frac{3}{2}-s-\epsilon\right)}{4^{s} \pi^{\frac{3}{2}-\epsilon}\Gamma(s)}\frac{1}{(x^{2})^{\frac{3}{2}-s-\epsilon}}.
\ee
In Landau gauge, for the gauge field propagators   we find
\be
\begin{split}
\langle (A_{\mu})_{i}^{\ j}(x)  (A_{\nu})_{k}^{\  l} (y)\rangle_{0}=&-\delta_{i}^{l}\delta_{k}^{j}\left(\frac{2\pi i}{\kappa}\right)\epsilon_{\mu\nu\rho}\partial^{\rho}_{x}\left(\frac{\Gamma\left(\frac{1}{2}-\epsilon\right)}{4 \pi^{\frac{3}{2}-\epsilon}}\frac{1}{((x-y)^{2})^{\frac{1}{2}-\epsilon}}\right),\\
\langle (\widehat A_{\mu})_{\hat i}^{\ \hat j}(x)  (\widehat A_{\nu})_{\hat k}^{\ \hat l} (y)\rangle_{0}=&\delta_{\hat i}^{\hat l}\delta^{\hat j}_{\hat k}\left(\frac{2\pi i}{\kappa}\right)\epsilon_{\mu\nu\rho}\partial^{\rho}_{x}\left(\frac{\Gamma\left(\frac{1}{2}-\epsilon\right)}{4 \pi^{\frac{3}{2}-\epsilon}}\frac{1}{((x-y)^{2})^{\frac{1}{2}-\epsilon}}\right).\\
\end{split}
\ee
The scalar propagators are instead given by
\be
\label{scal1}
\begin{split}
\langle (C_{I})_{i}^{\ \hat j}(x)  (\bar C^{J})_{\hat k}^{\  l} (y)\rangle_{0}=&\delta^{J}_{I}~\delta_{i}^{l}~\delta_{\hat k}^{\hat j}\frac{\Gamma\left(\frac{1}{2}-\epsilon\right)}{4 \pi^{\frac{3}{2}-\epsilon}}\frac{1}{((x-y)^{2})^{\frac{1}{2}-\epsilon}}\equiv  \delta^{J}_{I}~\delta_{i}^{l}~\delta_{\hat k}^{\hat j} D(x-y).
\end{split}
\ee
Finally we shall consider  the  case of the tree level  fermionic two-point function\footnote{Our choice of Dirac algebra is defined by $\gamma^{\mu}\gamma^{\nu}=\delta^{\mu\nu}-i\epsilon^{\mu\nu\rho}\gamma_{\rho}$}
\be
\label{fermionpropagator}
\left\langle
(\psi_{I})_{\hat i}^{\  j}(x)(\bar\psi^{J})_{ k}^{\ \hat l}(y)\right\rangle_{0}=\delta^{J}_{I}~\delta_{\hat i}^{\hat l}~\delta_{ k}^{ j} i\gamma^{\mu}\partial_{\mu}\left(\frac{\Gamma\left(\frac{1}{2}-\epsilon\right)}{4 \pi^{\frac{3}{2}-\epsilon}}\frac{1}{((x-y)^{2})^{\frac{1}{2}-\epsilon}}\right).
\ee
The  one-loop  corrections to the fermion propagator is given in by
\be
\label{ferm1}
\left\langle
(\psi_{I})_{\hat i}^{\  j}(p)(\bar\psi^{J})_{ k}^{\ \hat l}(-p)\right\rangle^{1~\rm \ell oop}_{0}=-2 i \delta_{\hat i }^{\\\hat l}\delta^{j}_{k}(N-M) \frac{16^{\epsilon -1} \pi ^{\epsilon } \left(p^2\right)^{-\frac{1}{2}-\epsilon } \sec (\pi  \epsilon )}{\Gamma (1-\epsilon )}.
\ee
Notice that this expression is finite when $\epsilon$ approaches zero. Its expression in coordinate space
is then obtained  by taking the Fourier-transform 
\be
\label{ferm2}
\left\langle
(\psi_{I})_{\hat i}^{\  j}(x)(\bar\psi^{J})_{ k}^{\ \hat l}(y)\right\rangle^{1~\rm \ell oop}_{0}=
-i \delta_{\hat i }^{\\\hat l}\delta^{j}_{k}(N-M)\frac{\Gamma ^{2}\left(\frac{1}{2}-\epsilon \right)}{16 \pi ^{3-2 \epsilon} }\frac{1}{((x-y)^{2})^{1-2\epsilon}}.
\ee
The last ingredient that is necessary for our analysis of  the two-loop behavior of the circle in  ABJ(M) theory is  the integral 
\be
\label{B9}
\begin{split}
\Gamma^{\lambda\mu\nu}(x_{1},x_{2}, x_{2})=&\left(\frac{\Gamma(\frac{1}{2}-\epsilon)}{4\pi^{3/2-\epsilon}}\right)^{3}\partial_{x_{1}^{\lambda}}\partial_{x_{2}^{\mu}}\partial_{x_{3}^{\nu}}
\int 
\frac{d^{3-2\epsilon}w}{(x_{1w}^{2})^{1/2-\epsilon}(x_{2w}^{2})^{1/2-\epsilon}(x_{3w}^{2})^{1/2-\epsilon}}\equiv\\
\equiv&\partial_{x_{1}^{\lambda}}\partial_{x_{2}^{\mu}}\partial_{x_{3}^{\nu}}\Phi,
\end{split}
\ee
which governs all the three-point functions appearing in our analysis.  The double contracted 3-point functions are then given by
\begin{subequations} 
\label{C10}
\begin{align}
\Gamma^{\lambda\lambda\rho}(x_{1},x_{2},x_{3})=&\partial_{x_{3}^{\rho}}(\partial_{x_{1}}\cdot \partial_{x_{2}})\Phi=
\frac{1}{2}
\partial_{x_{3}^{\rho}}[\square_{x_{3}}-\square_{x_{1}}-\square_{x_{2}}]\Phi\equiv
\partial_{x_{3}^{\rho}}\Phi_{3,12},\\
\Gamma^{\lambda\rho\lambda}(x_{1},x_{2},x_{3})
=&\partial_{x_{2}^{\rho}}(\partial_{x_{1}}\cdot \partial_{x_{3}})\Phi=\frac{1}{2}
\partial_{x_{2}^{\rho}}[\square_{x_{2}}-\square_{x_{1}}-\square_{x_{3}}]\Phi\equiv
\partial_{x_{2}^{\rho}}\Phi_{2,13},\\
\Gamma^{\rho\lambda\lambda}(x_{1},x_{2},x_{3})
=&\partial_{x_{1}^{\rho}}(\partial_{x_{2}}\cdot \partial_{x_{3}})\Phi=\frac{1}{2}
\partial_{x_{1}^{\rho}}[\square_{x_{1}}-\square_{x_{2}}-\square_{x_{3}}]\Phi\equiv
\partial_{x_{1}^{\rho}}\Phi_{1,23},
\end{align}
\end{subequations}
where  we took advantage of the invariance of  the scalar function $\Phi$ under translations [$(\partial_{x_{1}^{\lambda}}+\partial_{x_{2}^{\lambda}}+\partial_{x_{3}^{\lambda}})\Phi=0$]  and introduced the short-hand  notation
\be
\label{C11}
\begin{split}
\Phi_{i,jk}=& -\frac{\Gamma^{2}(1/2-\epsilon)}{32\pi^{3-2\epsilon}}\!\!
\left[\frac{1}{(x^{2}_{ij})^{\frac{1}{2}-\epsilon}(x^{2}_{ik})^{\frac{1}{2}-\epsilon}}-\frac{1}{(x^{2}_{ij})^{\frac{1}{2}-\epsilon}(x^{2}_{kj})^{\frac{1}{2}-\epsilon}}-\frac{1}{(x^{2}_{ik})^{\frac{1}{2}-\epsilon}(x^{2}_{jk})^{\frac{1}{2}-\epsilon}}\right]\!\!.
\end{split}
\ee 
In our computation we are also led to consider the value of  $\Phi_{i,jk}$ at coincident points. For $\epsilon>1/2$ they are finite and given by
\be
\Phi_{i,ik}= \frac{1}{2}\left(\frac{\Gamma(1/2-\epsilon)}{4\pi^{3/2-\epsilon}}\right)^{2}
\frac{1}{(x^{2}_{ik})^{1-2\epsilon}},\ \ \ \ \ \ \ \ \ \ \  
\Phi_{i,jj}= -\frac{1}{2}\left(\frac{\Gamma(1/2-\epsilon)}{4\pi^{3/2-\epsilon}}\right)^{2}
\frac{1}{(x^{2}_{ij})^{1-2\epsilon}}.
\ee 
In the spirit of DRED we extend these result to any value of $\epsilon$\footnote{
This is equivalent to the usual statement that massless tadpoles vanish in dimensional regularization.}.
\section{ Some properties of the function $F(\tau)$}
\label{Ftau}
The key ingredients in the evaluation of the double exchange diagram are the values in zero of the function 
\be
F(\tau)\equiv-\int^{2\pi}_{\tau}\!\!\! d\tau_{1}\int^{2\pi}_{\tau_{1}}\!\!\! d\tau_{2}\int^{2\pi}_{\tau_{2}}\!\!\! d\tau_{3}~\frac{1}{[\sin^{2}\frac{\tau_{3}}{2}]^{1-\epsilon}}=-\frac{1}{2}\int^{2\pi}_{\tau}\!\!\! d\tau_{1}
\frac{(\tau_{1}-\tau)^{2}}{[\sin^{2}\frac{\tau_{1}}{2}]^{1-\epsilon}},
\ee
and of its derivatives. The third derivative with respect to $\tau$,
\be
\partial_{\tau}^{3}F(\tau)=
\frac{1}{[\sin^{2}\frac{\tau}{2}]^{1-\epsilon}},
\ee
reproduces the building block appearing in the double exchange diagram.
By definition
$
F(2\pi)=F^{\prime}(2\pi)=F^{\prime\prime}(2\pi)=0.
$
Moreover this function  obeys a simple  reflection formula when $\tau\mapsto 2\pi -\tau$. In fact
\begin{align}
\label{B3}
F(2\pi-\tau)=&-\frac{1}{2}\int^{2\pi}_{2\pi-\tau}\!\!\! d\tau_{2}
\frac{(\tau_{2}-2\pi+\tau)^{2}}{[\sin^{2}\frac{\tau_{2}}{2}]^{1-\epsilon}}=-\frac{1}{2}\int^{\tau}_{0}\!\!\! d\tau_{2}
\frac{(\tau-\tau_{2})^{2}}{[\sin^{2}\frac{\tau_{2}}{2}]^{1-\epsilon}}=\nonumber\\
=&-\frac{1}{2}\int^{2\pi}_{0}\!\!\! d\tau_{2}
\frac{(\tau-\tau_{2})^{2}}{[\sin^{2}\frac{\tau_{2}}{2}]^{1-\epsilon}}- F(\tau)=F(0)+F^{\prime}(0) \tau+\frac{\tau^{2}}{2} F^{\prime\prime}(0)-F(\tau).
\end{align}
If we evaluate \eqref{B3} for $\tau=2\pi$ we obtain a relation connecting the first and the second derivative in $0$
\be
\label{PP}
F(0)=F(0)+2\pi F^{\prime}(0) +2\pi^{2} F^{\prime\prime}(0)-F(2\pi)\ \ \ \ \Rightarrow\ \ \ \  F^{\prime}(0)=-\pi F^{\prime\prime}(0).
\ee
We also have similar reflection formulae for the derivatives
\be
F^{\prime}(2\pi-\tau)= -F^{\prime}(0) -\tau F^{\prime\prime}(0)+F^{\prime}(\tau)\ \  \ \  \mathrm{and}\  \ \ \ F^{\prime\prime}(2\pi-\tau)=  F^{\prime\prime}(0)-F^{\prime\prime}(\tau).
\ee  
In order to determine the value of $F$ and its derivatives  in zero, we  need to find a more appropriate representation. To achieve this goal we shall consider   the following  identity 
\be
\frac{1}{[\sin^{2}\frac{\tau}{2}]^{1-\epsilon}}=\frac{2}{(2\epsilon-1)\epsilon} \frac{d^2}{d\tau^2}\left(\left(\sin^{2}\frac{\tau}{2}\right)^\epsilon\right)+\frac{2\epsilon}{(2\epsilon-1)}\left(\sin^{2}\frac{\tau}{2}\right)^\epsilon.
\ee
If we integrate three times both sides we obtain an alternative representation of  $F(\tau)$ which is better behaved around $\epsilon=0$,
 \be
 \label{F8}
 F(\tau)=\frac{2}{(1-2\epsilon)\epsilon}\int^{2\pi}_\tau d\tau^\prime \left(\sin^{2}\frac{\tau^\prime}{2}\right)^\epsilon+\frac{\epsilon}{(1-2\epsilon)}\int^{2\pi}_\tau d\tau^\prime (\tau-\tau^\prime)^2 \left(\sin^{2}\frac{\tau^\prime}{2}\right)^\epsilon.
  \ee
From this representation, we immediately find that 
\be
F(0) =\frac{4\pi}{\epsilon}+8\pi+2\int_{0}^{2\pi} \log\left[\sin^{2}\frac{\tau}{2}\right] +O(\epsilon)=\frac{4\pi}{\epsilon}+8\pi-8\pi\log 2+O(\epsilon).
\ee
We consider now the  second derivative in zero. By means of the representation \eqref{F8}, we can write
\begin{align}
 F^{\prime\prime}(0)=&\frac{2\epsilon}{(1-2\epsilon)}\int^{2\pi}_0 d\tau^\prime \left(\sin^{2}\frac{\tau^\prime}{2}\right)^\epsilon= 4\pi \epsilon+O(\epsilon^{2}).
\end{align}
The last step is to compute the only relevant quantity which is still an integral of $F$
\be
\int_{0}^{2\pi}\!\!\!\! d\tau  \left(2 \pi -\tau \right)
   (F^{\prime\prime}\left(\tau \right))^{2}.
\ee
From \eqref{F8} we can represent the second derivative of $F$ as follows
  \be
 F^{\prime\prime}(\tau)=-\frac{2}{(1-2\epsilon)\epsilon} \frac{d}{d\tau} \left(\sin^{2}\frac{\tau}{2}\right)^\epsilon+\frac{2\epsilon}{(1-2\epsilon)}\int^{2\pi}_\tau d\tau^\prime \left(\sin^{2}\frac{\tau^\prime}{2}\right)^\epsilon\equiv \frac{1}{\epsilon}g^{\prime\prime}(\tau)+\epsilon g(\tau).
  \ee
 and use this identity to evaluate the integral
 \begin{align}
   &\int_{0}^{2\pi}\!\!\!\! d\tau  \left(2 \pi -\tau \right)
   (F^{\prime\prime}\left(\tau \right))^{2}=\int_{0}^{2\pi}\!\!\!\! d\tau  \left(2 \pi -\tau \right)
\left (\frac{1}{\epsilon}g^{\prime\prime}(\tau)+\epsilon g(\tau)\right)^{2}=\nn\\
=&\frac{1}{\epsilon^{2}}\int_{0}^{2\pi}\!\!\!\! d\tau  \left(2 \pi -\tau \right)
\left (g^{\prime\prime}(\tau)\right)^{2}+2 \int_{0}^{2\pi}\!\!\!\! d\tau  \left(2 \pi -\tau \right) g^{\prime\prime}(\tau) g(\tau)+\epsilon^{2}\int_{0}^{2\pi}\!\!\!\! d\tau  \left(2 \pi -\tau \right) g^{2}(\tau)=\nn\\
=&-\frac{4}{(1-2 \epsilon )^2}\int_{0}^{2\pi}\!\!\!\! d\tau \left(2 \pi -\tau \right)
\left[\sin^2\left(\frac{\tau}{2}\right)\right]^{2 \epsilon}\!-\!\frac{4 F^{\prime}(0)}{(1-2 \epsilon )^2}-2 \int_{0}^{2\pi}\!\!\!\! d\tau  \left(2 \pi -\tau \right)( g^{\prime}(\tau))^{2} - g^{2}(0)+\nn\\
&+\epsilon^{2}\int_{0}^{2\pi}\!\!\!\! d\tau  \left(2 \pi -\tau \right) g^{2}(\tau)=-8\pi^{2}-16\pi^{2}-16\pi^{2}+O(\epsilon)=-40 \pi^{2}+O(\epsilon).
   \end{align}
\section{Vertex master integral}
\label{AppA}
The hardcore of the computation of  the  vertex diagram  discussed in subsec. \ref{Vertdiag}  is  governed by the 
tensor integral 
\be
\label{Vmn}
\begin{split}
\mathbf{V}^{\mu\nu}
=&-\left(\frac{\Gamma(\frac{3}{2}-\epsilon)}{2\pi^{3/2-\epsilon}}\right)^{3}
\int d^{3-2\epsilon}w
\frac{w^{\mu}w^{\nu}}{(x_{1w}^{2})^{3/2-\epsilon}(x_{2w}^{2})^{3/2-\epsilon}(x_{3w}^{2})^{3/2-\epsilon}},
\end{split}
\ee
saturated with the unit vector $h_{\mu}$ normal to the plane where our circle lies.  We find convenient  and efficient to  reduce  \eqref{Vmn} to scalar integrals  through the technique developed in \cite{Davydychev:1991va}, which will briefly summarize below for our specific example. The starting point is a scalar integral of the form
\begin{align}
J(n|\nu_{1},\nu_{2},\nu_{3})=&\frac{\Gamma(\nu_{1}) \Gamma(\nu_{2}) \Gamma(\nu_{3}) }{(2\pi^{\frac{n}{2}})^{3}}
\int\frac{ d^{n}w}{(x_{1w}^{2})^{\nu_{1}}(x_{2w}^{2})^{\nu_{2}}(x_{3w}^{2})^{\nu_{3}}}.
\end{align}
By introducing the standard Feynman parameters we can perform the integration over $w$  to obtain the following  representation for $ J(n|\nu_1,\nu_2,\nu_3)$
\begin{align}
\label{JJ}
J(n|\nu_1,\nu_2,\nu_3)
=&\frac{\Gamma\left (\nu_1+\nu_2+\nu_3-\frac{n}{2}\right)}{8\pi^{n}}\int_0^1dt_1 dt_2 d t_3 t_1^{\nu_1-1} t_2^{\nu_2-1}\ t_3^{\nu_3-1}\delta(1-t_1-t_2-t_3)\times\nn\\
&\times [ t_1 t_2 (x_{12})^2+t_1 t_3 (x_{13})^2+t_2 t_3 (x_{23})^2]^{\frac{n}{2}-\nu_1-\nu_2-\nu_3},
\end{align}
where we  have introduced for convenience the shorthand notation $x_{ij}^{\mu}= x_{i}^{
\mu}-x_{j}^{\mu}$.
If we take the derivative of  both sides  of \eqref{JJ} with respect to $x^{\mu}_{1}$ we find  
\begin{align}
&\partial_{x^{\mu}_{1}}J(n|\nu_1,\nu_2,\nu_3)
=-\frac{2\Gamma\left (\nu_1+\nu_2+\nu_3+1-\frac{n}{2}\right)}{8\pi^{n}}\int_0^1dt_1 dt_2 d t_3 t_1^{\nu_1-1} t_2^{\nu_2-1}\ t_3^{\nu_3-1}\times\\ &\times\!\delta(1\!-\!t_1\!-\!t_2\!-\!t_3)
 [ t_1 t_2 (x_{12})^2+t_1 t_3 (x_{13})^2+t_2 t_3 (x_{23})^2]^{\frac{n}{2}-\nu_1-\nu_2-\nu_3-1} ( t_1 t_2 x_{12}^\mu +t_1 t_3 x_{13}^{\mu}).\nn
\end{align}
The above expression can be  then rewritten in terms of scalar integral of the type  \eqref{JJ}, but in higher dimension:
\be
\partial_{x^{\mu}_{1}}J(n|\nu_1,\nu_2,\nu_3)\!=\!-2\pi^{2}[ x_{12\mu} J(n+2|\nu_{1}+1,\nu_{2}+1,\nu_{3})+
x_{13\mu} J(n+2|\nu_{1}+1,\nu_{2},\nu_{3}+1)].
\ee
We take now a derivative  of the above result with respect to $x_{2}^{\nu}$. We obtain
 \begin{align}
\partial_{x^{\mu}_{1}} \partial_{x^{\nu}_{2}}&J(n|\nu_1,\nu_2,\nu_3)\!=\!-2\pi^{2}[ -\hat\delta_{\mu\nu} J(n+2|\nu_{1}+1,\nu_{2}+1,\nu_{3})+ \nn\\
&+ x_{12\mu} \partial_{x^{\nu}_{2}}J(n+2|\nu_{1}+1,\nu_{2}+1,\nu_{3})+
x_{13\mu}\partial_{x^{\nu}_{2}} J(n+2|\nu_{1}+1,\nu_{2},\nu_{3}+1)],
\end{align}
where the symbol  $\hat\delta_{\mu\nu} $ denotes the  Kronecker delta in $3-2\epsilon$ dimensions. Since  the coordinates $x_{i}$ are  all orthogonal to  $h$, we can immediately write 
\be
\begin{split}
h_{\mu} h_{\nu} \mathbf{V}^{\mu\nu}=&-\frac{1}{4} h^{\mu} h^{\nu}\partial_{x^{\mu}_{1}} \partial_{x^{\nu}_{2}}J\left(3-2\epsilon\left|\frac{1}{2}\right.-\epsilon,\frac{1}{2}-\epsilon, \frac{3}{2}-\epsilon\right)=\\
=&\frac{\pi^{2}}{2} h^{\mu} h^{\nu}\hat \delta_{\mu\nu}J\left(5-2\epsilon\left|\frac{3}{2}\right.-\epsilon,\frac{3}{2}-\epsilon, \frac{3}{2}-\epsilon\right).
\end{split}
\ee
The computation of the contraction  $h^{\mu} h^{\nu}\hat \delta_{\mu\nu}$ requires particular care.  We first notice that the  bilinear   can be also rearranged as follows
\be
h^{\mu} h^{\nu}=\delta^{\mu\nu}- v_{1}^{\mu} v_{1}^{\nu}-v_{2}^{\mu} v_{2}^{\nu},
\ee
where  $\delta^{\mu\nu}$ is the  Kronecker delta in three dimensions and  $v_{i}$ are two orthonormal directions in plane where the circle  lies. Thus
\be
h^{\mu} h^{\nu}\hat \delta_{\mu\nu}=(\delta^{\mu\nu}- v_{1}^{\mu} v_{1}^{\nu}-v_{2}^{\mu} v_{2}^{\nu})
\hat \delta_{\mu\nu}=\delta^{\mu\nu} \hat \delta_{\mu\nu} -2=3-2\epsilon-2=(1-2\epsilon).
\ee
Here we used the DRED prescription that $\delta^{\mu\nu} \hat \delta_{\mu\nu}=3-2\epsilon$. 

\noindent
Exploiting the above result we obtain the following  higher dimensional representation for  our master  integral
\be
\begin{split}
h_{\mu} h_{\nu} \mathbf{V}^{\mu\nu}
=&\frac{\pi^{2}}{2} (1-2\epsilon) J\left(5-2\epsilon\left|\frac{3}{2}\right.-\epsilon,\frac{3}{2}-\epsilon, \frac{3}{2}-\epsilon\right).
\end{split}
\ee
\subsection{Asymptotic behavior}
\label{AppA1}
The second  key-ingredient of our analysis is the short distance behavior  [$x^\mu_{ij}\equiv x^{\mu}_i-x^{\mu}_j\to 0$] of  the vertex integral 
\begin{align}
 J\left(5-2\epsilon\left|\frac{3}{2}\right.-\epsilon,\frac{3}{2}-\epsilon, \frac{3}{2}-\epsilon\right).
\end{align}
This feature can be extracted from the  familiar representation of {\small $J\left(5-2\epsilon\left|\frac{3}{2}\right.-\epsilon,\frac{3}{2}-\epsilon, \frac{3}{2}-\epsilon\right)$} as an 
integral over Feynman parameters 
\begin{align}
\!\!\!
\frac{\Gamma(2-2\epsilon)}{8 \pi^{5-2\epsilon}}\int_0^1\!\!\!\! dt_1 dt_2 d t_3 ~\frac{t_1^{\frac{1}{2}-\epsilon} t_2^{\frac{1}{2}-\epsilon}\ t_3^{\frac{1}{2}-\epsilon}\delta(1\!-\!t_1\!-\!t_2\!-\!t_3)}{[ t_1 t_2 x_{12}^2+t_1 t_3 x_{13}^2+t_2 t_3 x_{23}^2]^{2-2\epsilon}},
\end{align}
by adapting the analysis performed  in app, B of  \cite{Alday:2010zy} to our case.  Since the integral is symmetric in the coordinates $x_i$ we can focus on the limit
$x_{12}^2\to 0$ without loss of generality  and perform the change of variable
\be
t_1=\frac{s}{1+z},\ \ t_2=\frac{1-s}{1+z},\ \  t_3 =\frac{z}{1+z} \ \  \ \  \mbox{where}\ \ \ \  0\le z\le\infty \  \mathrm{and} \   0\le s\le1.
\ee
We obtain
\be
\frac{\Gamma(2-2\epsilon)}{8 \pi^{5-2\epsilon}}
\int_0^\infty dz \int_0^1 ds
{\bar s}^{\frac{1}{2}-\epsilon } s^{\frac{1}{2}-\epsilon } z^{\frac{1}{2}-\epsilon } (z+1)^{-\epsilon -\frac{1}{2}} \left(\bar s s~
   x^{2}_{12}+z(\bar s  x^2_{23} +s x^2_{13})\right){}^{2 \epsilon -2},
\ee
where $\bar s=1-s$. In order to single out the singular behavior  we perform an additional change of variable,
namely we scale the $z$ as follows
\be
z\to \frac{{\bar s} s~ x^{2}_{12}~
   z}{{\bar s} x^{2}_{23}+s x^{2}_{13}}
\ee
and we get
\be
\label{A16}
\frac{\Gamma(2-2\epsilon)}{8 \pi^{5-2\epsilon}(x^{2}_{12})^{\frac{1}{2}-\epsilon }}
\int_0^\infty \frac{dz    z^{\frac{1}{2}-\epsilon }}{ (z+1)^{2-2 \epsilon}} \int_0^1 \!\!\! ds
   \left({\bar s} x^{2}_{23}+s
   x^2_{13}\right){}^{2 \epsilon -1}
   \left({\bar s} s x^{2}_{12}
   z+{\bar s} x^{2}_{23}+s
   x^{2}_{13}\right){}^{-\epsilon
   -\frac{1}{2}}.
\ee
The leading beaviour for $x^{2}_{12}\to 0$  can be now safely obtained since both the integrals over $z$ and over $s$ are convergent  when we pose  $x^{2}_{12}=0$ inside
\be
\begin{split}
&\frac{\Gamma(2-2\epsilon)}{8 \pi^{5-2\epsilon}}\frac{1}{(x^{2}_{12})^{\frac{1}{2}-\epsilon }}
\int_0^\infty \frac{dz    z^{\frac{1}{2}-\epsilon }}{ (z+1)^{2-2 \epsilon}} \int_0^1 \!\!\! ds
   \left({\bar s} x^{2}_{23}+s
   x^2_{13}\right){}^{ \epsilon -\frac{3}{2}}
   +O(1)=\\
=&   -\frac{
   \Gamma
   \left(\frac{1}{2}-\epsilon
   \right)^2}{8\pi ^{5-2 \epsilon}
  }   
  \frac{1}{(x^{2}_{12})^{\frac{1}{2}-\epsilon }} \left(\frac{1}{(x^{2}_{13})^{\frac{1}{2}-\epsilon}}-\frac{1}{(x^{2}_{23})^{\frac{1}{2}-\epsilon}}\right) \frac{1}{ \left(x^{2}_{13}-x^{2}_{23}\right)} +O(1).
   \end{split}
\ee
Since in this limit  $x_{13}^{2}$ becomes   $x_{23}^{2}$, we can  also write
\be
 J\left(5-2\epsilon\left|\frac{3}{2}\right.-\epsilon,\frac{3}{2}-\epsilon, \frac{3}{2}-\epsilon\right)
\simeq  \frac{(1-2 \epsilon)}{\pi^{2}} \left(\frac{
   \Gamma
   \left(\frac{1}{2}-\epsilon
   \right)}{4\pi ^{\frac{3}{2}- \epsilon}  }   \right)^{2}
  \frac{1}{(x^{2}_{12})^{\frac{1}{2}-\epsilon }} \frac{1}{(x^{2}_{23})^{\frac{3}{2}-\epsilon}}+O(1).
\ee
Therefore,  for the contraction $h_{\mu} h_{\nu} \mathbf{V}^{\mu\nu}$, we arrive to the following expansion 
\be
\label{Asbeh}
\begin{split}
h_{\mu} h_{\nu} \mathbf{V}^{\mu\nu}
=&\frac{(1-2\epsilon)^2}{2}  \left(\frac{
   \Gamma
   \left(\frac{1}{2}-\epsilon
   \right)}{4\pi ^{\frac{3}{2}- \epsilon}  }   \right)^{2}
  \frac{1}{(x^{2}_{12})^{\frac{1}{2}-\epsilon }} \frac{1}{(x^{2}_{23})^{\frac{3}{2}-\epsilon}}+O(1),
  \end{split}
\ee
 when $x_{12}^2$ approaches zero.  The behavior when $x_{13}^2$ or $x_{23}^2$ goes to zero is similar and it is simply
 obtained by permuting the role of the coordinates in \eqref{Asbeh}.

 \noindent
Consider now the  contraction $x_{1\mu} \Gamma^{\mu\rho\rho}$. If the  points $x_i$ belong to the unit circle,  we can greatly 
simplify its explicit form
\begin{align}
x_{1\mu} \Gamma^{\mu\rho\rho}
&\!=\!\frac{\left(1\!-\!2\epsilon\right)}{4}\! \!\left(\!\frac{\Gamma\left(\frac{1}{2}-\epsilon\right)}{4\pi^{3/2-\epsilon}}\right)^{\!\!2}\!\!\!\left(\!\frac{2}{(x_{12}^2)^{\frac{1}{2}-\epsilon}
(x_{13}^2)^{\frac{1}{2}-\epsilon}}\!-\!\frac{1}{(x^2_{12})^{\frac{1}{2}-\epsilon}
(x_{23}^2)^{\frac{1}{2}-\epsilon}}\!-\!\frac{1}{(x_{23}^2)^{\frac{1}{2}-\epsilon}
(x_{13}^2)^{\frac{1}{2}-\epsilon}}\!\!\right)\!\!.
\end{align}
Then the combination
\be
[(x_{12}^2+x_{13}^2)h_{\mu} h_{\nu} \mathbf{V}^{\mu\nu}-2 (1-2\epsilon) x_{1\mu}\Gamma^{\mu\rho\rho}]
\ee
is completely regular when either   $x_{12}^2$ or $x_{13}^2$ approach zero.

For $\epsilon\ne 0$ a closed  expression of  $h_{\mu} h_{\nu} \mathbf{V}^{\mu\nu}$  in terms of hypergeometric functions can be derived with help of the 
results given in \cite{Boos:1987bg}. However to achieve our goal it is  sufficient to know its value at $\epsilon=0$, which is obtained by directly performing the integral
\eqref{A16}
\be
\begin{split}
h_{\mu} h_{\nu} \mathbf{V}^{\mu\nu}
=&\frac{\pi^{2}}{2}  J\left(5\left|\frac{3}{2},\right.\frac{3}{2}, \frac{3}{2}\right)=
\frac{1}{16 \pi ^2 \sqrt{x_{12}^{2}}\sqrt{x_{13}^{2}}\sqrt{x_{23}^{2}}
   \left(\sqrt{x_{12}^{2}}+\sqrt{x_{13}^{2}}+\sqrt{x_{23}^{2}}\right)}.
\end{split}
\ee
\section{Integrals of contracted 3-point functions}
\label{AppD}
The goal of this appendix is to  provide more details on the procedure used to 
evaluate the integrals involving the double contracted three-point functions. 
We  first consider the following combination appearing in the vertex contribution
\begin{align}
\label{Cab1}
I_{1}=i\left(\frac{2\pi}{\kappa}\right)^{2}\!\!\!N M\!\!\int_{0}^{2\pi} \!\!\! \!\!\! d\tau_{1}\!\int_{\tau_{1}}^{2\pi}\!\!\! \!\!\! d\tau_{2}\!
\int_{\tau_{2}}^{2\pi}\!\!\! \!\!\! d\tau_{3}&\left[(\eta_{1}\bar\eta_{2}) \dot{x}_{3\lambda}(\Gamma^{\tau\lambda\tau}-\Gamma^{\lambda\tau\tau})+(\eta_{2}\bar\eta_{3}) \dot{x}_{1\nu}(\Gamma^{\tau\tau\nu}-\Gamma^{\tau\nu\tau})+\right.\nn\\
&+\left.(\eta_{1}\bar\eta_{3})\dot{x}_{2\nu}(\Gamma^{\tau\tau\nu}-\Gamma^{\nu\tau\tau})\right].
\end{align}
The integral  \eqref{Cab1} can be simplified if we use the symmetry of the integrand under the  exchange of the variable of integrations.  Note in fact it possesses the following structure
\be
\int_{0}^{2\pi}\!\!\!\!d\tau_{1}\int_{\tau_{1}}^{2\pi} \!\!\!\!d\tau_{2}\int_{\tau_{2}}^{
2\pi}\!\!\!\!d\tau_{3} ~[Q_{12,3}+ Q_{23;1}+Q_{13;2}]
\ee
where  $Q_{ij,k}=-Q_{ji,k}$. The subscripts on $Q$ summarizes its dependence on $\tau_{1}$, $\tau_{2}$ and $\tau_{3}$. Since $\tau_{3}>\tau_{2}>\tau_{1}$ we can rewrite  this integral as follows
\be
\int_{0}^{2\pi}\!\!\!\!d\tau_{1}\int_{\tau_{1}}^{2\pi} \!\!\!\!d\tau_{2}\int_{\tau_{2}}^{
2\pi}\!\!\!\!d\tau_{3} ~[Q_{12,3}~\sign(\tau_{2}-\tau_{1})+ Q_{23;1}~
\sign(\tau_{3}-\tau_{1})+Q_{13;2}~\sign(\tau_{3}-\tau_{1})].
\ee
In this form the integrand is manifestly symmetric in the exchange of all variable and thus  we can close the region of integration on the cube $[0,2\pi]\times [0,2\pi]\times [0,2\pi] $
\begin{align}
\label{D4}
&\!\frac{1}{3!}\int_{0}^{2\pi}\!\!\!\!\!\!d\tau_{1}\int_{0}^{2\pi} \!\!\!\!\!\!d\tau_{2}\int_{0}^{
2\pi}\!\!\!\!\!\!d\tau_{3} ~[Q_{12,3}~\sign(\tau_{2}-\tau_{1})+ Q_{23;1}~
\sign(\tau_{3}-\tau_{2})+Q_{13;2}~\sign(\tau_{3}-\tau_{1})]=\nn\\
=&\!\frac{1}{2}\int_{0}^{2\pi}\!\!\!\!\!\!d\tau_{1}\int_{0}^{2\pi} \!\!\!\!\!\!d\tau_{2}\int_{0}^{
2\pi}\!\!\!\!\!\!d\tau_{3} ~ Q_{23;1}~
\sign(\tau_{3}-\tau_{2})=\\
=&\!\frac{1}{2}\int_{0}^{2\pi}\!\!\!\!\!\!d\tau_{1}\int_{0}^{2\pi} \!\!\!\!\!\!d\tau_{2}\int_{\tau_{2}}^{
2\pi}\!\!\!\!\!\!d\tau_{3} ~ Q_{23;1}-
\frac{1}{2}\int_{0}^{2\pi}\!\!\!\!\!\!d\tau_{1}\int_{0}^{2\pi} \!\!\!\!\!\!d\tau_{3}\int_{\tau_{3}}^{
2\pi}\!\!\!\!\!\!d\tau_{2} ~ Q_{23;1}=\int_{0}^{2\pi}\!\!\!\!\!\!d\tau_{1}\int_{0}^{2\pi} \!\!\!\!\!\!d\tau_{2}\int_{\tau_{2}}^{
2\pi}\!\!\!\!\!\!d\tau_{3} ~ Q_{23;1}.\nn
\end{align}
If we apply this result to our original integral \eqref{Cab1} we get the following compact form
\begin{align}
\label{Cab2}
I_{1}=i\left(\frac{2\pi}{\kappa}\right)^{2}\!\!\!N M\!\!\int_{0}^{2\pi} \!\!\! \!\!\! d\tau_{1}\!\int_{\tau_{1}}^{2\pi}\!\!\! \!\!\! d\tau_{2}\!
\int_{\tau_{2}}^{2\pi}\!\!\! \!\!\! d\tau_{3}&~(\eta_{2}\bar\eta_{3}) \dot{x}_{1\nu}(\Gamma^{\tau\tau\nu}-\Gamma^{\tau\nu\tau}).
\end{align}
Next we put  the integrand   \eqref{Cab2}  into a form which is amenable to a direct  contour integration.  If we use the representation of the contracted three point functions 
in terms of  the auxiliary function $\Phi_{i,jk}$ defined in appendix \ref{ABJ}, we find
\begin{align}
\label{G1}
&(\eta_{2}\bar\eta_{3}) \dot{x}_{1\nu}(\Gamma^{\tau\tau\nu}-\Gamma^{\tau\nu\tau})=
(\eta_{2}\bar\eta_{3}) \dot{x}_{1}\cdot(\partial_{x_{3}}\Phi_{3,12}-\partial_{x_2}\Phi_{2,13})=\\
&=(\eta_{2}\bar\eta_{3}) \dot{x}_{1}\cdot(\partial_{x_{3}}(\Phi_{3,12}+\Phi_{2,13})+\partial_{x_1}\Phi_{2,13})
=(\eta_{2}\bar\eta_{3}) \left(\dot{x}_{1}\cdot\partial_{x_{3}}(\Phi_{3,12}+\Phi_{2,13})+\frac{d}{d\tau_{1}}\Phi_{2,13}\right),\nn
\end{align}
where we used the invariance under translation of the function $\Phi_{i,jk}$: $(\partial_{x_{1}}+\partial_{x_{2}}+\partial_{x_{3}})\Phi_{i,jk}=0$. Exploiting the fact that the combination $x^{2}_{ij}$ on a circle is a function only of $\tau_{i}-\tau_{j}$, the first term  in \eqref{G1} can be rewritten as follows
\begin{align}
&(\eta_{2}\bar\eta_{3}) \dot{x}_{1}\cdot\partial_{x_{3}}(\Phi_{3,12}+\Phi_{2,13})=-(\eta_{2}\bar\eta_{3})\left(\frac{\Gamma(1/2-\epsilon)}{4\pi^{3/2-\epsilon}}\right)^{2}
\frac{1}{(x_{12}^{2})^{1/2-\epsilon} }\frac{d}{d\tau_{1}}\frac{1}{ (x^{2}_{13})^{1/2-\epsilon}}=\nn\\
&=(\eta_{2}\bar\eta_{3})\left(\frac{\Gamma(1/2-\epsilon)}{4\pi^{3/2-\epsilon}}\right)^{2}
\frac{1}{(x_{12}^{2})^{1/2-\epsilon} }\frac{d}{d\tau_{3}}\frac{1}{ (x^{2}_{13})^{1/2-\epsilon}}=\nn\\
&=\frac{d}{d\tau_{3}} [(\eta_{2}\bar\eta_{3})  (\Phi_{3,12}+\Phi_{2,13})]-
(\Phi_{3,12}+\Phi_{2,13})\frac{d}{d\tau_{3}}(\eta_{2}\bar\eta_{3}) .
\end{align}
Therefore we have to compute the  contour integral 
\begin{align}
\label{DEWa}
I_{1}&=i\left(\frac{2\pi}{\kappa}\right)^{2}\!\!\!N M\!\!\int_{0}^{2\pi} \!\!\! \!\!\! d\tau_{1}\!\int_{0}^{2\pi}\!\!\! \!\!\! d\tau_{2}\!
\int_{\tau_{2}}^{2\pi}\!\!\! \!\!\! d\tau_3 \biggl[\frac{d}{d\tau_{3}}[(\eta_{2}\bar\eta_{3})  (\Phi_{3,12}+\Phi_{2,13})]-\nn\\
&-(\Phi_{3,12}+\Phi_{2,13})\frac{d}{d\tau_{3}}(\eta_{2}\bar\eta_{3}) +\frac{d}{d\tau_{1}}[(\eta_{2}\bar\eta_{3}) \Phi_{2,13}]\biggr].
\end{align}
The last term in \eqref{DEWa} is a total derivative in $\tau_{1}$  of a periodic function of this variable.   The integral over the whole period is  then zero. We remain with 
\begin{align}
\label{DEWa1}
I_{1}&=i\left(\frac{2\pi}{\kappa}\right)^{2}\!\!\!N M\!\!\int_{0}^{2\pi} \!\!\! \!\!\! d\tau_{1}\!\int_{0}^{2\pi}\!\!\! \!\!\! d\tau_{2}\!
\int_{\tau_{2}}^{2\pi}\!\!\! \!\!\! d\tau_3 \biggl[\frac{d}{d\tau_{3}}[(\eta_{2}\bar\eta_{3})  (\Phi_{3,12}+\Phi_{2,13})]-(\Phi_{3,12}+\Phi_{2,13})\frac{d}{d\tau_{3}}(\eta_{2}\bar\eta_{3}) \biggr].
\end{align}
We consider first the term which is a total derivative with respect to $\tau_{3}$, and we perform the integration over $\tau_{3}$
\begin{align}
&\int_{0}^{2\pi} \!\!\! \!\!\! d\tau_{1}\!\int_{0}^{2\pi}\!\!\! \!\!\! d\tau_{2}\!
\int_{\tau_{2}}^{2\pi}\!\!\! \!\!\! d\tau_3 \frac{d}{d\tau_{3}}[(\eta_{2}\bar\eta_{3})  (\Phi_{3,12}+\Phi_{2,13})]=\int_{0}^{2\pi} \!\!\! \!\!\! d\tau_{1}\!\int_{0}^{2\pi}\!\!\! \!\!\! d\tau_{2}\Bigl[ [ (\eta_{2}\bar\eta_{3})  (\Phi_{3,12}+\Phi_{2,13})]_{\tau_{3}=2\pi}-\nn\\
&-2i (\Phi_{2,12}+\Phi_{2,12})\Bigr] =-2i\left(\frac{\Gamma(1/2-\epsilon)}{4\pi^{3/2-\epsilon}}\right)^{2}\!\!\!\!\int_{0}^{2\pi}\!\!\! \!\!\! d\tau_{1}\!
\int_{0}^{2\pi}\!\!\! \!\!\! d\tau_2\biggr[ \frac{\cos\frac{\tau_{2}}{2}}{(4 \sin ^2\left(\frac{\tau _1-\tau
   _2}{2}
\right))^{\frac{1}{2}-\epsilon} (4 \sin ^2\left(\frac{\tau _1}{2}
\right))^{\frac{1}{2}-\epsilon}}+\nn\\
&+\frac{2}{ (4 \sin ^2\left(\frac{\tau _1-\tau
   _2}{2}
\right))^{1-2\epsilon}}\biggr].\end{align}
The first term gives a vanishing integral since  is odd in the transformation $(\tau_{1},\tau_{2})\mapsto (2\pi -\tau_{1},2\pi-\tau_{2})$. The second term is  proportional to the integral appearing one-loop and thus vanishes when $\epsilon \to 0$. 

The evaluation of $I_{1}$ collapses to
\begin{align}
I_{1}&=-\left(\frac{2\pi}{\kappa}\right)^{2}\!\!\!N M\!\!\int_{0}^{2\pi} \!\!\! \!\!\! d\tau_{1}\!\int_{0}^{2\pi}\!\!\! \!\!\! d\tau_{2}\!
\int_{\tau_{2}}^{2\pi}\!\!\! \!\!\! d\tau_3 \sin\frac{\tau_{3}-\tau_{2}}{2} (\Phi_{3,12}+\Phi_{2,13})=\nn\\
&=-\left(\frac{2\pi}{\kappa}\right)^{2}\!\!\!N M\left(\frac{\Gamma(1/2-\epsilon)}{4\pi^{3/2-\epsilon}}\right)^{2}\int_{0}^{2\pi} \!\!\! \!\!\! d\tau_{1}\!\int_{0}^{2\pi}\!\!\! \!\!\! d\tau_{2}\!
\int_{\tau_{2}}^{2\pi}\!\!\! \!\!\! d\tau_3 
\frac{\sin\frac{\tau_{3}-\tau_{2}}{2} }{(4 \sin ^2\left(\frac{\tau _1-\tau
   _2}{2}
\right))^{\frac{1}{2}-\epsilon}(4 \sin ^2\left(\frac{\tau _1-\tau
   _3}{2}
\right))^{\frac{1}{2}-\epsilon}}.
\end{align}
The integration over the contour  becomes  straightforward  once we observe that the integrand is the sum of two total derivatives, one w.r.t. $\tau_{2}$ and one  w.r.t. $\tau_{3}$, if  we write
\be
\sin\frac{\tau_{3}-\tau_{2}}{2} =\sin\frac{\tau_{3}-\tau_{1}}{2} \cos\frac{\tau_{1}-\tau_{2}}{2}+ \sin\frac{\tau_{1}-\tau_{2}}{2} \cos\frac{\tau_{3}-\tau_{1}}{2}. 
\ee 
This allows us to  perform easily one of the integrals  and to remain with two closed
integration
\be 
I_{1}=-\frac{1}{\kappa^{2}\epsilon} 4^{2 \epsilon -2} \pi ^{2
   \epsilon -1} \Gamma
   \left(\frac{1}{2}-\epsilon
   \right)^2
\!\!\!N M\!\!\int_{0}^{2\pi} \!\!\! \!\!\! d\tau_{1}\!\int_{0}^{2\pi}\!\!\! \!\!\! d\tau_{2}
\left[ \sin ^2\left(\frac{\tau _1-\tau
   _2}{2}
   \right)\right]^{2 \epsilon
   }
  \ee
Since the integrand  is  a function of $\tau_{1}-\tau_{2}$,  it is  periodic and the integration is over the whole period,  the integral over $\tau_{1}$   yields a result independent of $\tau_{2}$. Therefore we can drop the integral over $\tau_{2}$ and
multiply by $2\pi$, we get
\begin{align}
I_{1}=&-\frac{2\pi}{\kappa^{2}\epsilon} 4^{2 \epsilon -2} \pi ^{2
   \epsilon -1} \Gamma
   \left(\frac{1}{2}-\epsilon
   \right)^2
\!\!\!N M\!\!\int_{0}^{2\pi} \!\!\! \!\!\! d\tau_{1}\!\left[ \sin ^2\left(\frac{\tau _1}{2}
   \right)\right]^{2 \epsilon
   }=\nn\\
   =&-\frac{2\pi}{\kappa^{2}\epsilon} 4^{2 \epsilon -2} \pi ^{2
   \epsilon -1} \Gamma
   \left(\frac{1}{2}-\epsilon
   \right)^2
\!\!\!N M\frac{2 \sqrt{\pi } \Gamma \left(2
   \epsilon
   +\frac{1}{2}\right)}{\Gamma (2
   \epsilon +1)}=\nn\\
=&-\frac{M N 4^{2 \epsilon -1} \pi ^{2 \epsilon +\frac{1}{2}} \Gamma \left(\frac{1}{2}-\epsilon \right)^2 \Gamma \left(2 \epsilon +\frac{1}{2}\right)}{\kappa ^2 \epsilon ^2 \Gamma (2 \epsilon )}=\nn\\
=&-\frac{\pi ^{2+2\epsilon} M N }{2 \kappa ^2  }\left(\frac{1}{\epsilon} +2 \gamma   +  4\log 2+O(\epsilon)\right).
\end{align}
Next we consider  the second type of contribution containing contracted three-point functions
\be
\mbox{\small$\displaystyle
I_{2}=4i(1-\epsilon) \left(\frac{2\pi}{\kappa}\right)^{2}\!\!\!N M\!\!\int_{0}^{2\pi} \!\!\! \!\!\! d\tau_{1}\!\int_{\tau_{1}}^{2\pi}\!\!\! \!\!\! d\tau_{2}\!
\int_{\tau_{2}}^{2\pi}\!\!\! \!\!\! d\tau_{3}\!\left[\frac{r_{12}}{(\eta_{2}\bar\eta_{1})} x_{3\nu}\Gamma^{\tau\tau\nu}+\frac{ r_{23}}{(\eta_{3}\bar\eta_{2})}x_{1\nu}\Gamma^{\nu\tau\tau}+\frac{r_{13}}{(\eta_{3}\bar\eta_{1})} x_{2\nu}\Gamma^{\tau\nu\tau}\right].$}
\ee
Using again the result \eqref{D4}, we can  reduce the integrand just one term. The contour integration can be performed along the same line discussed above  and one gets
\begin{align}
\label{V2a}
I_{2}=&4i(1-\epsilon) \left(\frac{2\pi}{\kappa}\right)^{2}\!\!\!N M\!\!\int_{0}^{2\pi} \!\!\! \!\!\! d\tau_{1}\!\int_{0}^{2\pi}\!\!\! \!\!\! d\tau_{2}\!
\int_{\tau_{2}}^{2\pi}\!\!\! \!\!\! d\tau_{3} \frac{ r_{23}}{(\eta_{3}\bar\eta_{2})}x_{1\nu}\Gamma^{\nu\tau\tau}=\nn\\
=&-\frac{M N 4^{2 \epsilon -1} \pi ^{2
   \epsilon } \Gamma
   \left(\frac{1}{2}-\epsilon
   \right)^2 \left(\pi  \epsilon 
   \Gamma (2 \epsilon )-\sqrt{\pi }
   \Gamma \left(2 \epsilon
   +\frac{1}{2}\right)\right)}{\kappa
   ^2 \epsilon ^2 \Gamma (2 \epsilon
   -2)}=\\
   =&-\frac{\pi ^{2+2\epsilon} M N }{2 \kappa ^2 
   }\left(\!-\frac{1}{\epsilon} +3-2 \gamma+O(\epsilon)\!\right)\!.\nn
\end{align}


\newpage

\begin{thebibliography}{99}
\bibitem{Aharony:2008ug}
  O.~Aharony, O.~Bergman, D.~L.~Jafferis and J.~Maldacena,
  ``{\it N=6 super-conformal  Chern- Simons matter theories, M2-branes and their gravity duals,}''
  JHEP {\bf 0810} (2008) 091
  [arXiv:0806.1218 [hep-th]]. 
  \bibitem{Aharony:2008gk}
  O.~Aharony, O.~Bergman and D.~L.~Jafferis,
  ``{\it Fractional M2-branes},''
  JHEP {\bf 0811} (2008) 043
  [arXiv:0807.4924 [hep-th]].   
\bibitem{Henn:2010ps}
  J.~M.~Henn, J.~ Plefka and K.~Wiegandt,
  ``{\it Light-like polygonal Wilson loops in 3d Chern- Simons and ABJM theory},''
  JHEP {\bf 1008} (2010) 032   
   \bibitem{Chen:2011vv}
  W.~-M.~Chen and Y.~-t.~Huang,
  {\it ``Dualities for Loop Amplitudes of N=6 Chern-Simons Matter Theory,''}
  JHEP {\bf 1111} (2011) 057
  [arXiv:1107.2710 [hep-th]].  
  \bibitem{Bianchi:2011dg}
  M.~S.~Bianchi, M.~Leoni, A.~Mauri, S.~Penati and A.~Santambrogio,
 {\it  ``Scattering Amplitudes/Wilson Loop Duality In ABJM Theory,''}
  JHEP {\bf 1201} (2012) 056
  [arXiv:1107.3139 [hep-th]].   
\bibitem{Wiegandt:2011uu} 
  K.~Wiegandt,
{\it ``Equivalence of Wilson Loops in ${\cal N} = 6$ super Chern-Simons matter theory and ${\cal N} = 4$ SYM Theory,''}
  Phys.\ Rev.\ D {\bf 84}, 126015 (2011)
  [arXiv:1110.1373 [hep-th]].
\bibitem{Bianchi:2013pva} 
  M.~S.~Bianchi, G.~Giribet, M.~Leoni and S.~Penati,
 {\it ``Light-like Wilson loops in ABJM and maximal transcendentality,''}
  arXiv:1304.6085 [hep-th].
\bibitem{Bianchi:2013iha} 
  M.~S.~Bianchi, M.~Leoni, M.~Leoni, A.~Mauri, S.~Penati and A.~Santambrogio,
 {\it ``ABJM amplitudes and WL at finite N,''}
  arXiv:1306.3243 [hep-th].
\bibitem{Drukker:2008zx}
  N.~Drukker, J.~Plefka  and D.~Young,
  {\it ``Wilson loops in 3-dimensional N=6 supersymmetric Chern-Simons Theory and
  their string theory duals,''}
  JHEP {\bf 0811}, 019 (2008)
  [arXiv:0809.2787 [hep-th]].
\bibitem{Chen:2008bp}
  B.~Chen and J.~-B.~Wu,
 {\it ``Supersymmetric Wilson Loops in N=6 Super Chern-Simons-matter theory,''}
  Nucl.\ Phys.\ B {\bf 825} (2010) 38
  [arXiv:0809.2863 [hep-th]].
\bibitem{Rey:2008bh}
  S.~J.~Rey, T.~Suyama and S.~Yamaguchi,
  {\it ``Wilson Loops in Superconformal Chern-Simons Theory and Fundamental Strings
  in Anti-de Sitter Supergravity Dual,''}
  JHEP {\bf 0903}, 127 (2009)
  [arXiv:0809.3786 [hep-th]].
\bibitem{Rey:1998ik}
  W.~J.~Rey and J.~T.~Yee,
  {\it  ``Macroscopic strings as heavy quarks in large N gauge theory and anti-de
  Witter supergravity,''}
  Eur.\ Phys.\ J.\  C {\bf 22}, 379 (2001)
  [arXiv:hep-th/9803001].
\bibitem{Maldacena:1998im}
  J.~M.~Maldacena,
  { \it ``Wilson loops in large N field theories,''}
  Phys.\ Rev.\ Lett.\  {\bf 80}, 4859 (1998)
  [arXiv:hep-th/9803002].  
\bibitem{Erickson:2000af}
  J.~K.~Erickson, G.~W.~Semenoff and K.~Zarembo,
{\it   ``Wilson loops in N=4 supersymmetric Yang-Mills theory,''}
  Nucl.\ Phys.\ B {\bf 582} (2000) 155
  [hep-th/0003055].       
\bibitem{Drukker:2000rr}
  N.~Drukker and D.~J.~Gross,
{\it   ``An Exact prediction of N=4 SUSYM for string theory,''}
  J.\ Math.\ Phys.\  {\bf 42}, 2896 (2001)   
\bibitem{Pestun:2007rz}
  V.~Pestun,
  {\it  ``Localization of gauge theory on a four-sphere and supersymmetric Wilson loops,''}
  arXiv:0712.2824 [hep-th].
\bibitem{Drukker:2007qr}
  N.~Drukker, S.~Giombi, R.~Ricci and D.~Trancanelli,
 { \it ``Supersymmetric Wilson loops on $S^3$,''}
  JHEP {\bf 0805}, 017 (2008)
  [arXiv:0711.3226 [hep-th]].    
\bibitem{Bassetto:2008yf}
  A.~Bassetto, L.~Griguolo, F.~Pucci and D.~Seminara,
  {\it ``Supersymmetric Wilson loops at two loops,''}
  JHEP {\bf 0806}, 083 (2008)
  [arXiv:0804.3973 [hep-th]].   
\bibitem{Young:2008ed}
  D.~Young,
  {\it ``BPS Wilson Loops on $S^2$ at Higher Loops,''}
  JHEP {\bf 0805}, 077 (2008)
  [arXiv:0804.4098 [hep-th]];
  JHEP {\bf 0908}, 061 (2009)
  [arXiv:0905.1943 [hep-th]].
\bibitem{Bassetto:2009rt} 
  A.~Bassetto, L.~Griguolo, F.~Pucci, D.~Seminara, S.~Thambyahpillai and D.~Young,  {\it ``Correlators of supersymmetric Wilson-loops, protected operators and matrix models in N=4 SYM,''}
  JHEP {\bf 0908}, 061 (2009)
  [arXiv:0905.1943 [hep-th]];  {\it ``Correlators of supersymmetric Wilson loops at weak and strong coupling,''}
  JHEP {\bf 1003}, 038 (2010)
  [arXiv:0912.5440 [hep-th]].
  \bibitem{Giombi:2012ep}
  S.~Giombi and V.~Pestun, {\it ``Correlators of local operators and 1/8 BPS Wilson loops on S$^{2}$ from 2d YM and matrix models,''}
  JHEP {\bf 1010}, 033 (2010)
  [arXiv:0906.1572 [hep-th]];
  {\it ``Correlators of Wilson Loops and Local Operators from Multi-Matrix Models and Strings in AdS,''}
  JHEP {\bf 1301} (2013) 101
  [arXiv:1207.7083 [hep-th]].
\bibitem{Dymarsky:2009si}
  A.~Dymarsky and V.~Pestun,
 { \it ``Supersymmetric Wilson loops in N=4 SYM and pure spinors,''}
  JHEP {\bf 1004}, 115 (2010)
  \bibitem{Cardinali:2012sy} 
  V.~Cardinali, L.~Griguolo and D.~Seminara,
 {\it  ``Impure Aspects of Supersymmetric Wilson Loops,''}
  JHEP {\bf 1206}, 167 (2012)
  [arXiv:1202.6393 [hep-th]].
\bibitem{Correa:2012at} 
  D.~Correa, J.~Henn, J.~Maldacena and A.~Sever,
  {\it ``An exact formula for the radiation of a moving quark in N=4 super Yang Mills,''}
  arXiv:1202.4455 [hep-th].
   \bibitem{Gromov:2012eu} 
  N.~Gromov and A.~Sever,
 {\it  ``Analytic Solution of Bremsstrahlung TBA,''}
  JHEP {\bf 1211}, 075 (2012)
  [arXiv:1207.5489 [hep-th]].
\bibitem{Gromov:2013qga} 
  N.~Gromov, F.~Levkovich-Maslyuk and G.~Sizov,
 {\it  ``Analytic Solution of Bremsstrahlung TBA II: Turning on the Sphere Angle,''}
  arXiv:1305.1944 [hep-th].
\bibitem{Drukker:2011za} 
  N.~Drukker and V.~Forini,
  {\it  ``Generalized quark-antiquark potential at weak and strong coupling,''}
  JHEP\ {\bf 1106}, 131  (2011)
  [arXiv:1105.5144 [hep-th]].
  \bibitem{Drukker:2012de} 
  N.~Drukker,
  {\it ``Integrable Wilson loops,''}
  arXiv:1203.1617 [hep-th].
  \bibitem{Correa:2012hh} 
  D.~Correa, J.~Maldacena and A.~Sever,
 {\it ``The quark anti-quark potential and the cusp anomalous dimension from a TBA equation,''} JHEP {\bf 1208}, 134 (2012)
  [arXiv:1203.1913 [hep-th]].
\bibitem{Kapustin:2009kz}
  A.~Kapustin, B.~Willett and I.~Yaakov,
{\it ``Exact Results for Wilson Loops in Superconformal Chern-Simons Theories with Matter,''}
  JHEP {\bf 1003} (2010) 089
  [arXiv:0909.4559 [hep-th]].
\bibitem{Marino:2009jd}
  M.~Marino and P.~Putrov,
{\it ``Exact Results in ABJM Theory from Topological Strings,''}
  JHEP {\bf 1006} (2010) 011
  [arXiv:0912.3074 [hep-th]].
\bibitem{Drukker:2010nc}
  N.~Drukker, M.~Marino and P.~Putrov,
 {\it ``From weak to strong coupling in ABJM theory,''}
  Commun.\ Math.\ Phys.\  {\bf 306} (2011) 511
  [arXiv:1007.3837 [hep-th]].
\bibitem{Drukker:2011zy}
  N.~Drukker, M.~Marino and P.~Putrov,
 {\it ``Nonperturbative aspects of ABJM theory,''}
  JHEP {\bf 1111} (2011) 141
  [arXiv:1103.4844 [hep-th]].
\bibitem{Drukker:2009hy} 
  N.~Drukker and D.~Trancanelli,   {\it ``A Supermatrix model for N=6 super Chern-Simons-matter theory,''}
  JHEP {\bf 1002}, 058 (2010)  [arXiv:0912.3006 [hep-th]].
 \bibitem{Berenstein:2008dc} 
  D.~Berenstein and D.~Trancanelli,
  {\it ``Three-dimensional N=6 SCFT's and their membrane dynamics,''}
  Phys.\ Rev.\ D {\bf 78}, 106009 (2008)
  [arXiv:0808.2503 [hep-th]].
\bibitem{Lee:2010hk}
  K.~M.~Lee and S.~Lee,
  {\it ``1/2-BPS Wilson Loops and Vortices in ABJM Model,''}
  JHEP {\bf 1009}, 004 (2010)
  [arXiv:1006.5589 [hep-th]].
\bibitem{Marino:2012az} 
  M.~Marino and P.~Putrov,
{\it   ``Interacting fermions and N=2 Chern-Simons-matter theories,''}
  arXiv:1206.6346 [hep-th].
\bibitem{Klemm:2012ii} 
  A.~Klemm, M.~Marino, M.~Schiereck and M.~Soroush,
 {\it  ``ABJM Wilson loops in the Fermi gas approach,''}
  arXiv:1207.0611 [hep-th].
\bibitem{Hatsuda:2012hm} 
  Y.~Hatsuda, S.~Moriyama and K.~Okuyama,
  {\it ``Exact Results on the ABJM Fermi Gas,''}
  JHEP {\bf 1210}, 020 (2012)
  [arXiv:1207.4283 [hep-th]].
\bibitem{Hatsuda:2013yua} 
  Y.~Hatsuda, M.~Honda, S.~Moriyama and K.~Okuyama,
 {\it  ``ABJM Wilson Loops in Arbitrary Representations,''}
 [arXiv:1306.4297 [hep-th]].
\bibitem{Marino:2011eh}
  M.~Marino and P.~Putrov,
{\it  ``ABJM theory as a Fermi gas,''}
  J.\ Stat.\ Mech.\  {\bf 1203} (2012) P03001
  [arXiv:1110.4066 [hep-th]].
\bibitem{Bianchi:2013zda} 
  M.~S.~Bianchi, G.~Giribet, M.~Leoni and S.~Penati,
 {\it  ``The 1/2 BPS Wilson loop in ABJM theory at two loops,''}
  arXiv:1303.6939 [hep-th]  to appear on Phys. Rev. D.
\bibitem{Griguolo:2012iq} 
  L.~Griguolo, D.~Marmiroli, G.~Martelloni and D.~Seminara,
{\it  ``The generalized cusp in ABJ(M) N = 6 Super Chern-Simons theories,''}
  Phys.\ Lett.\ B {\bf 718}, 615 (2012)
 [arXiv:1208.5766 [hep-th]].
 \bibitem{Witten:1988hf}
E.~Witten,
{\it   ``Quantum Field Theory and the Jones Polynomial,''}
  Commun.\ Math.\ Phys.\  {\bf 121} (1989) 351.
\bibitem{Siegel:1979wq} 
  W.~Siegel,
  {\it ``Supersymmetric Dimensional Regularization via Dimensional Reduction,''}
  Phys.\ Lett.\ B {\bf 84}, 193 (1979).
\bibitem{Penati2}
 M.~S.~Bianchi, G.~Giribet, M.~Leoni and S.~Penati, {\it `` The 1/2 BPS Wilson loop in ABJ(M) at two loops: The details''}, to appear on the arXiv concurrently, arXiv:1307.0786 [hep-th].
\bibitem{Guadagnini:1989am} 
  E.~Guadagnini, M.~Martellini and M.~Mintchev,
  {\it ``Wilson Lines in Chern-Simons Theory and Link Invariants,''}
  Nucl.\ Phys.\ B {\bf 330}, 575 (1990).
  \bibitem{Cardinali:2012ru} 
  V.~Cardinali, L.~Griguolo, G.~Martelloni and D.~Seminara,
  {\it ``New supersymmetric Wilson loops in ABJ(M) theories,''}
  Phys.\ Lett.\ B {\bf 718}, 615 (2012)
  [arXiv:1209.4032 [hep-th]].
\bibitem{Zarembo:2002an}
  K.~Zarembo,
{\it  ``Supersymmetric Wilson loops,''}
  Nucl.\ Phys.\ B {\bf 643} (2002) 157
  [hep-th/0205160].
\bibitem{Chen:1992ee}
 W.~Chen, G.~W.~Semenoff and Y.~S.~Wu,
 {\it ``Two loop analysis of nonAbelian Chern-Simons theory,''}
 Phys.\ Rev.\  D {\bf 46} (1992) 5521
 [hep-th/9209005].
 \bibitem{Benna:2008zy}
 M.~Benna, I.~Klebanov, T.~Klose and M.~Smedback,
 {\it ``Superconformal Chern-Simons Theories and AdS${}_4$/CFT${}_3$ Correspondence,''}
 arXiv:0806.1519  [hep-th].
  \bibitem{Davydychev:1991va} 
  A.~I.~Davydychev,
 {\it ``A Simple formula for reducing Feynman diagrams to scalar integrals,''}
  Phys.\ Lett.\ B {\bf 263}, 107 (1991).
  \bibitem{Alday:2010zy} 
  L.~F.~Alday, B.~Eden, G.~P.~Korchemsky, J.~Maldacena and E.~Sokatchev,
  {\it ``From correlation functions to Wilson loops,''}
  JHEP {\bf 1109}, 123 (2011)
  [arXiv:1007.3243 [hep-th]].
\bibitem{Boos:1987bg} 
  E.~E.~Boos and A.~I.~Davydychev,
  {\it ``A Method Of The Evaluation Of The Vertex Type Feynman Integrals,''}
  Moscow Univ.\ Phys.\ Bull.\  {\bf 42N3}, 6 (1987)
  [Vestn.\ Mosk.\ Univ.\ Fiz.\ Astron.\  {\bf 28N3}, 8 (1987)].

   




\end{thebibliography}
\end{document}